\documentclass[fleqn,usenatbib]{mnras}

\usepackage[T1]{fontenc}
\usepackage{ae,aecompl}

\usepackage[usenames]{color}
\usepackage{graphicx}
\usepackage{amsmath}
\usepackage{amssymb}
\usepackage[utf8]{inputenc}

\usepackage{txfonts}

\usepackage{gensymb}
\usepackage{subcaption}
\newcommand{\PAph}{$\text{PA}_{ph}$}
\newcommand{\PAkn}{$\text{PA}_{kn}$}
\newcommand{\CR}{$\mathrm{R}_{CR}$}
\newcommand{\Rbar}{$\mathrm{R}_{bar}$}
\newcommand{\Rdep}{$\mathrm{R}_{bar}^{dep}$} 
\newcommand{\Omph}{$\Omega_{bar}^{ph}$}
\newcommand{\Omkn}{$\Omega_{bar}^{kn}$}
\newcommand{\R}{$\mathcal{R}$}
\newcommand{\Om}{$\Omega_{bar}$}
\newcommand{\dOmp}{$\delta \Omega_{bar}^+$}
\newcommand{\dOmm}{$\delta \Omega_{bar}^-$}
\newcommand{\Xint}{$\langle X \rangle$}
\newcommand{\Vint}{$\langle V \rangle$}

\newcommand{\angcur}{$\omega (R)$}
\newcommand{\dPA}{ $\delta \Omega_{PA}$}
\newcommand{\dPAp}{$\delta \Omega_{PA}^{+}$}
\newcommand{\dPAm}{ $\delta \Omega_{PA}^{-}$}
\newcommand{\dC}{ $\delta \Omega_{C}$}
\newcommand{\dCp}{ $\delta \Omega_{C}^{+}$}
\newcommand{\dCm}{ $\delta \Omega_{C}^{-}$}
\newcommand{\dL}{ $\delta \Omega_{L}$}
\newcommand{\dLp}{ $\delta \Omega_{L}^{+}$}
\newcommand{\dLm}{ $\delta \Omega_{L}^{-}$}
\def\code#1{\texttt{#1}}
\def\Spear#1{$r_s={#1}$}

\title[Bar pattern speed in MaNGA galaxies]{SDSS-IV MaNGA: Bar pattern speed estimates with the Tremaine-Weinberg method and their error sources }

\author [Garma-Oehmichen et al.]{L. Garma-Oehmichen,$^{1}$ \thanks{E-mail: lgarma@astro.unam.mx} 
M. Cano-Díaz, $^{2}$
H. Hernández-Toledo, $^{1}$
E. Aquino-Ortíz, $^{1}$
\newauthor
O.Valenzuela, $^{1}$ 
J. A. L. Aguerri, $^{3}$ 
S. F. Sánchez $^{1}$
and M. Merrifield $^{4}$
\\
$^{1}$ Instituto de Astronomía, Universidad Nacional Autónoma de México, Apartado Postal 70-264, Mexico D.F., 04510, Mexico \\
$^{2}$CONACYT Research Fellow - Instituto de Astronomía, Universidad Nacional Autónoma de México, Apartado Postal 70-264, Mexico D.F., 04510, Mexico \\
$^{3}$ Instituto de Astrofísica de Canarias. C/ Vía Láctea s/n, 38200 La Laguna, Spain \\
$^{4}$ School of Physics \& Astronomy, University of Nottingham, University Park, Nottingham, NG7 2RD, U.K.}

\date{Accepted XXX. Received YYY; in original form ZZZ}

\pubyear{2019}

\begin{document}
\label{firstpage}
\pagerange{\pageref{firstpage}--\pageref{lastpage}}
\maketitle

% Abstract of the paper
\begin{abstract}
    Estimating the bar pattern speed (\Om{}) is one of the main challenges faced in understanding the role of stellar bars in galaxy dynamical evolution. This work aims to characterise different uncertainty sources affecting the Tremaine Weinberg (TW)-method to study the correlation between bar and galaxies physical parameters. We use a sample of 15 MaNGA SDSS-IV galaxies and 3 CALIFA galaxies from \cite{Aguerri2015}. We studied the errors related with (i) galaxy centre determination, (ii) disc position angle (PA) emphasising the difficulties triggered by outer non-axisymmetric structures besides the bar, (iii) the slits length and (iv) the spatial resolution. In average, the PA uncertainties range $\sim 15 \%$, the slit length $\sim 9 \%$ and the centring error $\sim 5 \%$. Reducing the spatial resolution increases the sensitivity to the PA error. Through Monte Carlo simulations, we estimate the probability distribution of the \R{} bar speed parameter. The present sample is composed of 7 slow, 4 fast and 7 ultrafast bars, with no trend with morphological types. Although uncertainties and low sample numbers may mask potential correlations between physical properties, we present a discussion of them: We observe an anti-correlation of \Om{} with the bar length and the stellar mass, suggesting that massive galaxies tend to host longer and slower bars. We also observe a correlation of the molecular gas fraction with \R{}, and a weak anti-correlation with \Om{}, suggesting that bars rotate slower in gaseous discs. Confirmation of such trends awaits future studies.
\end{abstract}


%________________________________________________________________

\section{Introduction}

The presence of a bar highly influences the galaxy dynamics and secular evolution, by redistributing angular momentum, energy and mass between different components, such as the disk, the bulge and the dark matter halo \citep{Weinberg1985, Debattista1998,Valenzuela2003, Athanassoula2013, Kormendy2004, Sellwood2014}. This exchange of angular momentum happens mainly by the bar resonances, transporting angular momentum from the inner region to outside corrotation \citep{LyndenBell1972, Tremaine1984b, Athanassoula2003}. Bars can induce a substantial amount of gas inflow, triggering star formation in the central regions \citep{Hernquist1995, Martinent1997} possibly fueling the active galactic nucleus \citep{Laine2002}, while quenching the inner kiloparsec region \citep{Gavazzi2015} and the rest of the galaxy \citep{Carles2016, Khoperskov2018}. Barred galaxies also tend to show flatten abundance profiles \citep{Seidel2016, Fraser2019}, resulting from the gaseous flows \citep{Kubryk2015} and stellar radial migration \citep{Kubryk2013, DiMatteo2013}. The presence of a bar could also influence the properties of other structures like resonance rings \citep{Schwarz1981, Buta1996, Rautiainen2000} and spiral arms \citep{Block2004, Salo2010, Dobbs2014, Romero2006, Romero2007}.

Almost one-third of nearby galaxies host bars larger than 4 kpc \citep{Sellwood1993, Menendez2007, Aguerri2009}. Small nuclear bars are usually hidden behind dust, so they are best seen in near-infrared bands where the bar fraction can increase to $\sim 70 \%$ \citep{Eskridge2000, Whyte2002, Marinova2007}. The bar fraction is strongly dependent on galaxy mass (or luminosity) \citep{Nair2010,  MendezAbreu2012, Erwin2018}. It is not clear if this fraction remains constant \citep{Jogee2004, Elmegreen2004} or decreases towards higher redshifts \citep{Sheth2008, Melvin2014}. 

The bar pattern speed \Om{} is one of the most important parameters to understand how barred galaxies evolve, since it determines the location of the resonances and the rate of angular momentum exchange. In principle, \Om{} has a physical upper limit. Studies of stellar orbits in barred potentials show that self-consistent bars cannot extend outside the corotation resonance radius (hereafter \CR{}) since the stellar orbits become elongated perpendicular to the bar. Besides, the increasing density of resonances near corotation leads to chaos in phase space \citep{Contopoulos1980}. The pattern speed is often parametrized with the dimensionless parameter $\mathcal{R} = \mathrm{R}_{CR}/\mathrm{R}_{bar}$. Since \CR{} is the natural upper bound for \Rbar{}, bars with \R{} close to unity rotate as fast as nature allows. By convention, bars are defined as ``fast'' if $\mathcal{R} < 1.4$ and ``slow'' if  $\mathcal{R} > 1.4$. Given the difficulties for measuring \Rbar{} and \Om{}, estimates of \R{} are plagued by uncertainties. However, the vast majority of galaxies appear to be fast \citep[e.g.,][]{Rautiainen2008, Corsini2011, Aguerri2015}.

Several model-dependent methods have been developed to determine $\Omega_{bar}$. Hydrodynamical simulations of individual galaxies can recover $\Omega_{bar}$ by matching the modelled and observed gas distribution and/or gas velocity field \citep{Sanders1980, Hunter1988, Lindblad1996, Aguerri2001, Rautiainen2008}. The gravitational torque produced by the rotating bar causes gas flowing in the radial direction and accumulating in major resonances where the net torque vanishes \citep{Buta1996}, leading to methods that rely on the location of morphological features such as spiral arms \citep{Puerari1997, Aguerri1998}, rings \citep{Schwarz1981, Rautiainen2000, Patsis2003}, or leading dust lanes \citep{Athanassoula1992, Sanchez-Menguiano2015}. Another approach is to measure directly the corotation radius by looking for a change of sign in the streaming motions of the gas \citep{Font2011, Font2017},  locating the dark gaps in ringed galaxies \citep{Buta2017} and the shift between potential and density waves patterns \citep{Zhang2007, Buta2009}.

Based on the continuity equation, the only model-independent method for estimating \Om{} is the so-called \cite{Tremaine1984} (hereafter TW) method.  Using the stellar light as a tracer, the TW-method has been mostly used in early-type barred galaxies which are not obscured by the dust and have no star formation \citep{Kent1987, Merrifield1995, Gerssen1999, Debattista2002, Aguerri2003, Corsini2003, Debattista2004, Corsini2007}. The general applicability of the TW-method has allowed other tracers to be used such as the stellar mass distribution \citep{Gerssen2007, Aguerri2015, Guo2018}, gas tracers such as CO \citep{Zimmer2004, Rand2004} and H$\alpha$ \citep{Hernandez2005, Emsellem2006, Fathi2007, Chemin2009, Gabbasov2009, Fathi2009}.

The arrival of large galaxy surveys based on the Integral Field Spectroscopic (IFS) technique, has extended the applicability of the TW-method to larger samples of galaxies \citep{Aguerri2015, Guo2018, Cuomo2019, Zou2019}. Efforts are being made to see if there are correlations between \Om{} and other galactic properties (or, equally interestingly, if there are not). However, the ability to measure such correlations depends critically on the errors on each determination.

The sensitivity of the TW-method to the Position Angle (PA), slits length and spatial resolution has been well documented for simulated galaxies \citep{Debattista2003, Guo2018, Zou2019}. However, a characterization of these errors based on observations of real galaxies has not yet been done. In this work, we carefully quantify PA, slits lengths and centring errors in real galaxies, establishing a methodology to determine these errors for future measurements. 
In this paper, we use a pilot sample of 15 MaNGA galaxies and 3 CALIFA galaxies from \cite{Aguerri2015} to study these errors. With a robust estimate of errors, this sample may be useful to look for preliminary correlations with other galactic properties.

The paper is organized as follows: In Section \ref{sec:TW_method} we explain in detail the TW-method and its natural limitations. In Section \ref{sec:Data} we present the data used and our sample selection. In Section \ref{Sec:Parameters} a description of the photometric and kinematic analysis is presented. In Section \ref{sec:Errors} the nature of  different error sources is described, and the $\Omega_{bar}$ estimates are presented. In Section \ref{Sec:Discussion} we present a general discussion on the  error sources and possible constrains. Finally in Section \ref{Sec:Conclusions} we summarize our results and give our conclusions. Throughout this paper we use a Hubble constant $H_0=73$ km s$^{-1}$ Mpc$^{-1}$ \citep{Riess2018}.
\section{The Tremaine-Weinberg method}
\label{sec:TW_method}

The TW-method requires a tracer that satisfies the continuity equation, like the stellar population within a galaxy or the mass distribution. Assuming the disc of the galaxy is flat and has a well-defined pattern speed the TW-equation is:

\begin{equation}
\Omega_{bar}\sin i = \frac{\int_{-\infty}^{\infty} h(Y) \int_{-\infty}^{\infty} \Sigma(X,Y)V_{\parallel}(X,Y) \mathrm{d}X \mathrm{d}Y } {\int_{-\infty}^{\infty} h(Y)  \int_{-\infty}^{\infty} X \Sigma (X,Y) \mathrm{d}X \mathrm{d}Y }
\label{Eq:TW}
\end{equation}

where $i$ is the galaxy inclination, $(X, Y)$ are the Cartesian coordinates in the sky plane, with the origin at the galaxy centre, and the $X$-axis aligned with the line of nodes. $\Sigma(X, Y)$ and  $V_{\parallel}(X, Y)$ are the surface brightness and the line of sight (LOS) velocity of the tracer respectively. $h(Y)$ is an arbitrary weight function, that is usually given by a delta function $\delta(Y - Y_0)$, that corresponds to a slit in long-slit spectroscopy. For integral-field spectroscopy, we can use the same weight function and refer to it as a pseudo-slit.

\cite{Merrifield1995}, noted that the integrals in equation \ref{Eq:TW} could be interpreted as luminosity-weighted means of the LOS velocity and position (hereafter \Vint{}  and \Xint{} respectively). In this sense, the TW-integrals can be calculated in any reference frame, particularly in the galaxy centre rest frame. Plotting both integrals for several pseudo-slits produces a straight line with slope $\Omega_{bar} \sin i$. 

\begin{equation}
\Omega_{bar}\sin i = \frac{\langle V \rangle}{\langle X \rangle}
\label{Eq:TW+}
\end{equation}
\begin{equation}
\langle V \rangle=\frac{\int_{-\infty}^{\infty} h(Y) \mathrm{d}Y \int_{-\infty}^{\infty} \Sigma(X,Y)(V_{\parallel}-V_{sys})(X,Y) \mathrm{d}X}{\int_{-\infty}^{\infty} h(Y) \mathrm{d}Y \int _{-\infty}^{\infty} \Sigma(X,Y)\mathrm{d}X}
\label{Eq:kinematic}
\end{equation} 
\begin{equation}
\langle X \rangle=\frac{\int_{-\infty}^{\infty} h(Y) \mathrm{d}Y \int_{-\infty}^{\infty} X \Sigma(X,Y)\mathrm{d}X}{\int_{-\infty}^{\infty} h(Y) \mathrm{d}Y \int_{-\infty}^{\infty} \Sigma(X,Y)\mathrm{d}X}
\label{Eq:photometric}
\end{equation} 

Nevertheless, the TW-method has some limitations and cannot be applied to all barred galaxies. In face-on galaxies, the kinematic information tends to be poor, while in edge-on galaxies the photometric information is lost. The orientation of the bar should also be taken into account. If the bar is oriented towards the galaxy's minor or major axis both \Xint{} and \Vint{} tend to cancel out. 

The symmetry of the integrals \ref{Eq:kinematic} and \ref{Eq:photometric}, means that all the axisymmetric contributions are canceled. Thus, the integration limits can be changed to $\pm X_{max}$ and $\pm Y_{max}$ if the axisymmetric disc is reached. Note, however, that the axisymmetric disc cancels out only if the slits are correctly aligned with the disc PA.

\begin{figure}
  \centering
  \includegraphics[width = \linewidth]{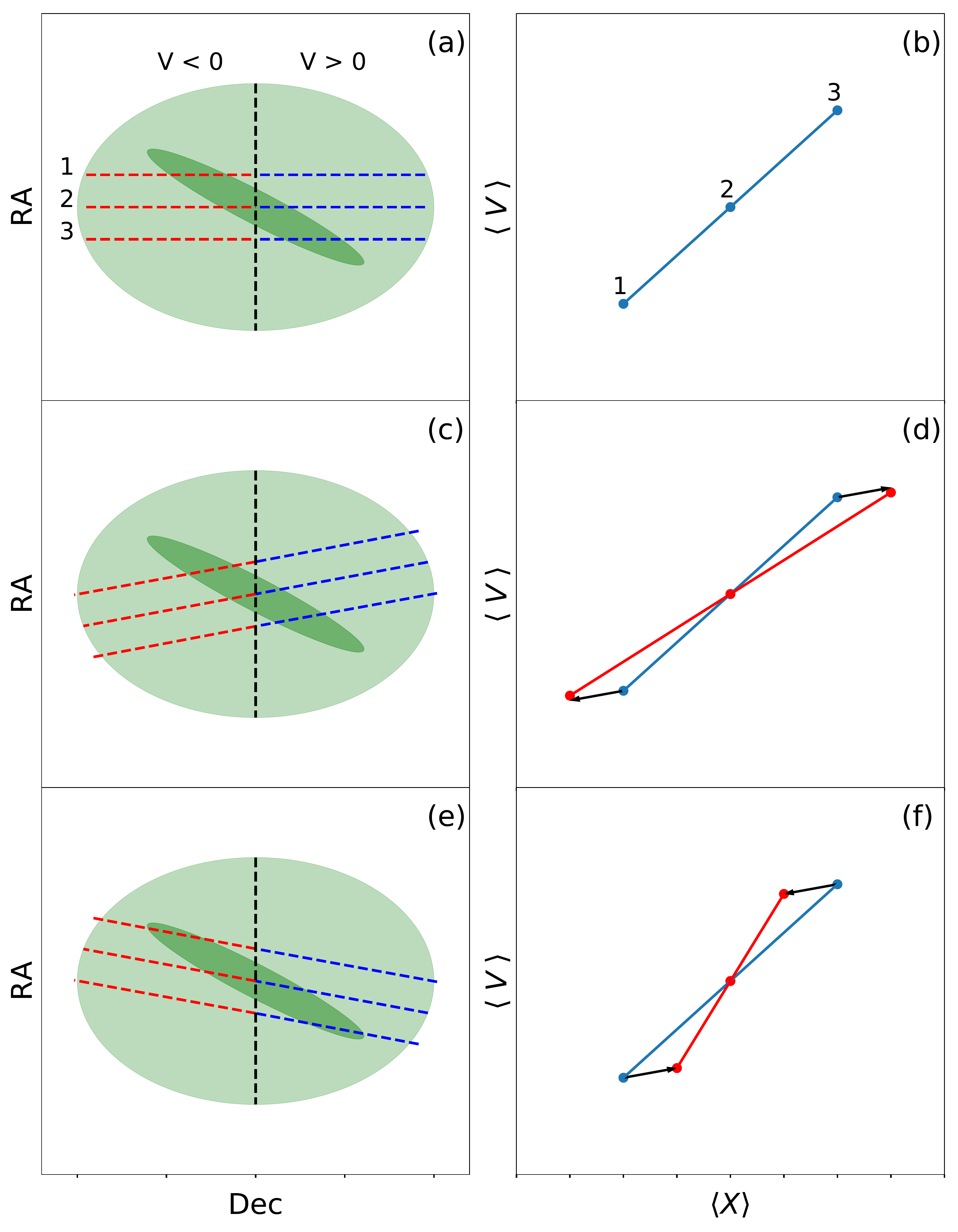} 
  \caption{The PA error in an ideal inclined barred galaxy. Depending on the orientation of the bar and the velocity field, an error in the PA measurement will change the values of \Vint{} and \Xint{} producing a fictitious pattern speed.}
  \label{Fig:TW_example}
\end{figure}

The sensitivity of the TW-method to the PA of the disc was first described by \cite{Debattista2003}, showing that in N-body simulations an error of 5\degree can produce errors as big as $48 \%$ in \Om{}. Figure \ref{Fig:TW_example} shows a simplified example of the method being applied to an inclined barred galaxy at its rest frame. The LOS velocity is chosen to be negative on the left side and positive on the right side. In panel (a) the pseudo-slits are aligned correctly with the disc PA, and when plotting \Xint{} vs \Vint{} in panel (b) we measure the correct \Om{}. In the middle and lower panels, the pseudo-slits are not aligned correctly and the axisymmetric light from the disc does not cancel out. Depending on the geometry of the velocity field, the orientation and shape of the bar, the values of \Vint{} and \Xint{} can increase or decrease producing a fictitious pattern speed. In this simplified picture, we only consider the change in \Xint{}. In panels (c) and (d) the new orientation increases the value  \Xint{} in the outer slits, producing a slightly lower value of \Om{}. In contrast, in panels (e) and (d) \Xint{} decreases, producing a dramatic increase in \Om{}. In general, the PA error is not symmetrical and is highly galaxy dependent. For this reason, a second PA from the galaxy kinematics is highly desirable, as it could help to constrain the errors.

Another important error source is the length of the slits. \cite{Guo2018} studied this error using a simulated barred galaxy from \cite{Athanassoula2013}. They observe that the pattern speed increases with the length of the slits, and flattens at \Om{} for slits longer than 1.2 times the bar length, as they explain in their Appendix A2. The associated error increases with the inclination angle and the PA difference between the disc and the bar. Also, their test shows that using slits shorter than the bar appears to underestimate the correct pattern.

The geometric nature of the TW-method leaves opens the question of whether or not other important sources of error still need to be fully considered. In this work, we characterize the PA, slit length, centring error as well as the resolution of the bar due to the PSF. We discuss the test we performed in more detail in Section \ref{sec:Errors}.
\section{Data}
\label{sec:Data}

\subsection{The MaNGA survey}

The Mapping Nearby Galaxies at Apache Point Observatory (MaNGA) survey is one of the three core programs of the Sloan Digital Sky Survey IV (SDSS - IV) Collaboration \citep{Blanton2017}. MaNGA in the process of acquiring IFS data for $\sim$ 10,000 galaxies in the local universe $(0.01 < z < 0.15)$ spanning all environments, morphologies and within a stellar mass range of $10^9 \: - \: 10^{11} M_\odot$ \citep{Bundy2015}. It uses the Baryonic Oscillation Spectroscopic Survey (BOSS) spectrograph, which provides a spectral resolution of $R \approx 2000$ $(\sigma_{inst} \approx 77$ km s$^{-1})$ in the wavelength range $3600 - 10300$ \r{A} \citep{Smee2013}. Target galaxies are chosen so that the final sample has a flat distribution in stellar mass, and uniform spatial coverage in units of effective radius $\mathrm{R}_e$. To accomplish the science goals, about $2/3$ of the sample is covered out to $1.5 \: \mathrm{R}_e$ (Primary sample), and $1/3$ to $2.5 \mathrm{R}_e$ (Secondary sample) \citep{Yan2016}.

The observations are performed using previously drilled plates, in which a set of 17 hexagonal fiber-bundle IFU's is plugged in. The diameter of each fiber is 2 arcsec. The number of fibers in each bundle ranges from 19 to 127 \citep{Drory2015}. A 3-point dithering pattern is adopted to have total coverage of the field of view \citep{Law2016}.

We will refer to the galaxies in this work, using the MaNGA identification (ID) number, which consists of the prefix 'manga', followed by two sets of numbers (each separated by the '-' symbol). The first one represents the plate used in the observation, while the second one serves to identify the number of fibers used to observe the target (first two to three digits) and to distinguish between targets observed with the same configuration within the same plate (last two digits). For example, the target 'manga-7495-12704' in our sample, was observed with the Plate ID 7495, using a bundle of 127 fibers, and was the object 04 to be observed with that configuration in this plate. 

\subsection{The CALIFA survey}

The details of the Calar Alto Legacy Integral Field Area (CALIFA) survey, including the observational strategy and data reduction, are explained in \citet{Sanchez2012} and \citet{DR3}. All galaxies were observed using PMAS \citep{Roth2005} in the PPaK configuration \citep{Kelz2006}, covering a hexagonal field of view (FoV) of 74$\arcsec$ $\times$64$\arcsec$, which is sufficient to map the full optical extent of the galaxies up to two to three disk effective radii. This is possible because of the diameter selection of the CALIFA sample \citep{Walcher2014}.

The observing strategy guarantees complete coverage of the FoV, with a final spatial resolution of FWHM$\sim$2.5$\arcsec$, corresponding to $\sim$1 kpc at the average redshift of the survey \citep[e.g][]{rgb15, DR3}. The sampled wavelength range and spectroscopic resolution for the adopted setup (3745 -- 7500 \AA, $\lambda/\Delta\lambda\sim$850, V500 setup) are more than sufficient to explore the most prominent ionized gas emission lines from [Oii]$\lambda$3727 to [Sii]$\lambda$6731 at the redshift of our targets, on one hand, and to deblend and subtract the underlying stellar population, on the other \citep[e.g.,][]{Kehrig2012, CidFernandes2013, CidFernandes2014, sanchez13, sanchez14, PipeI}.

The current dataset was reduced using version 2.2 of the CALIFA pipeline, whose modifications with respect the previous ones \citep{Sanchez2012, Husemann2013, rgb15} are described in \citet{DR3}. The final dataproduct after the reduction is a datacube comprising the spatial information in the {\it x} and {\it y} axis, and the spectral one in the {\it z} one. For further details of the adopted dataformat and the quality of the data consult \citet{DR3}.

\subsection{Sample selection}

The sample was selected from the first MaNGA public data release SDSS Data Release 13 \citep{Albareti2017}. A visual morphological classification (Vazquez-Mata et al. in prep) originally based on the SDSS images and later verified upon the DESI Legacy Imaging Surveys \citep{Dey2019} was used. 

Since the main purpose of this paper is to study the impact of various uncertainty sources on the estimate of \Om{}, we have selected galaxies under various conditions. This includes a few galaxies whose bar PA was relatively close to the PA disc axis (manga-7990-12704, manga-8243-12704), galaxies with weak bars (manga-8453-12701), galaxies with possible minor interactions (manga-8135-6103, manga-8453-12701, manga-8313-9101) and galaxies with low and high inclinations (manga-8341-12704 and manga-8317-12704 respectively). This results in a cutoff in inclination and PA difference between the disc and bar of $\sim 15 \degree$ and $\sim 6 \degree$ respectively.
From an original number of 200 galaxies under the above circumstances, we considered a final number of 15 MaNGA galaxies with good signal to noise ratio (S/N $\sim 10$) at the outermost regions of the IFU field of view. 

To compare our results and spot systematic errors in our methodology, we re-analysed a sub-sample of 3 CALIFA galaxies already analyzed with the TW-method by \cite{Aguerri2015} (NGC 5205, NGC 5406 and NGC 6497). According to their light-weighted measurements, these three galaxies host ultra-fast bars. In their mass-weighted results, only NGC 5406 becomes a slow bar. We choose these galaxies due to the controversy surrounding the existence of ultra-fast bars. In Section \ref{Sec:Discussion} we also compare and discuss the galaxies we have in common with \citep{Guo2018}. 

To facilitate the discussion in the following sections, we will use three galaxies from our sample, as key examples. These galaxies were chosen to illustrate how the measurement of the disc PA can be affected in different scenarios using a photometric or kinematic method (hereafter \PAph{} and \PAkn{}). (1) manga-8439-6102, a galaxy where both \PAph{} and \PAkn{} are similar and has one of the smallest errors in our sample. This is the same galaxy that is presented in \cite{Guo2018}. (2) manga-8135-6103, a strong barred galaxy where \PAph{} is biased due to the strong spiral arms, however, the kinematic orientation gives a better estimation of \Om{}. Also illustrates one case where the bar covers entirety the FoV of MaNGA. (3) manga-8341-12704, a strongly barred, and low inclined galaxy where the velocity map is highly perturbed, making the \PAkn{} more difficult to estimate. 

\subsection{Stellar flux and velocity maps}
\label{Sec:Pipe3D}

Along this article we use the intensity flux and velocity maps provided by Pipe3D for both, the MaNGA and CALIFA data sets. \code{Pipe3D} is a data analysis pipeline developed to analyze and characterize the stellar populations and the ionized gas of galaxies observed with the IFS technique \citep{PipeI,  PipeII}. \code{Pipe3D} uses a binning algorithm that is aimed to reach a goal in S/N while adding together areas of the galaxy with similar physical properties (like spiral arms, bars, interarm regions, etc.) by following the surface brightness pattern of the galaxies \citep[for more details on the binning see for example:][]{Sanchez2019}. The stellar properties in each bin are derived by fitting the co-added spectra to a set of single stellar population (SSP) templates, using the \code{FIT3D} fitting tool \citep{PipeI}. 

The resulting dataproducts are a set of maps for various physical or observational properties, such as emission lines fluxes, stellar masses, stellar and gaseous velocity maps, star formation histories, etc. For a detailed description of the procedure, including the binning algorithm, dust attenuation, uncertainties and S/N distribution please refer to \cite{PipeI, PipeII, Ibarra2016}.

In this work, we compute \Xint{} and \Vint{} by summing directly over the stellar flux and stellar velocity maps derived by \code{Pipe3D}. In Figure \ref{Fig:Rot_curves} we show these pair of maps for our example galaxies. We also make use of the H$_\alpha$ velocity maps to model the rotation curve (see Section \ref{Sec:Rotation_curve}).

In Table \ref{Tab:Sample} we present the main parameters of our sample, including the stellar mass and the molecular gaseous mass derived from the \code{Pipe3D} analysis. We should note that our molecular gas mass was estimated adopting the dust-to-gas ratio based on the dust attenuation obtained by Pipe3D. The current calibrator was already used in \cite{Sanchez2018} and \cite{Galbany2017}, and explained in detail in Barrera-Ballesteros et al (in prep.).

\begin{table*}
\caption{Main parameters of our sample}
\label{Tab:Sample}
\centering
{
\renewcommand{\arraystretch}{1.3}
\begin{tabular}{cccccccc}
\hline \hline
 Galaxy & RA & Dec & Morph & Distance  & $r_{50}$ & Stellar Mass & Gaseous Mass \\
        & (hh:mm:ss) & (hh:mm:ss) & & [Mpc] & [kpc] & $[\log \, M/M_\odot]$  & $[\log \, M/M_\odot]$ \\
  (1) & (2) & (3) & (4) & (5) & (6) & (7)  & (8) \\               
\hline 
 manga-7495-12704 &  13:41:45 &  27:00:16 &  SBbc &     118.6 &  7.1 &  10.7 &      10.1 \\
 manga-7962-12703 &  17:24:52 &  28:04:42 &  SBbc &     197.9 & 13.4 &  11.1 &      10.6 \\
 manga-7990-12704 &  17:29:57 &  58:23:51 &   SBa &     113.2 &  8.6 &  10.6 &       - \\
  manga-8135-6103 &  07:32:14 &  39:33:36 &  SBab &     201.3 &  8.9 &  10.9 &       9.7 \\
 manga-8243-12704 &  08:44:40 &  53:57:04 &  SBbc &      99.6 &  5.2 &  10.3 &       9.9 \\
  manga-8256-6101 &  10:54:56 &  41:29:54 &   SBb &     101.4 &  3.9 &  10.3 &       9.3 \\
  manga-8257-3703 &  11:06:37 &  46:02:20 &   SBb &     103.3 &  2.7 &  10.5 &       9.9 \\
 manga-8312-12704 &  16:29:13 &  41:09:03 &   SBa &     123.5 &  5.9 &  10.3 &       9.9 \\
  manga-8313-9101 &  15:58:47 &  41:56:17 &  SBbc &     160.7 &  6.5 &  10.9 &      10.3 \\
 manga-8317-12704 &  12:54:49 &  44:09:20 &  SBab &     223.8 & 19.1 &  11.1 &      10.6 \\
 manga-8318-12703 &  13:04:56 &  47:30:13 &  SBbc &     162.2 & 10.3 &  11.0 &      10.4 \\
 manga-8341-12704 &  12:36:51 &  45:39:04 &  SBbc &     125.3 &  5.7 &  10.7 &       9.9 \\
  manga-8439-6102 &  09:31:07 &  49:04:47 &   SBb &     140.2 &  5.1 &  10.8 &      10.1 \\
 manga-8439-12702 &  09:26:09 &  49:18:37 &  SBab &     111.5 &  9.5 &  10.6 &      10.1 \\
 manga-8453-12701 &  10:05:14 &  46:39:03 &  SABc &     103.7 &  6.0 &  10.3 &       9.9 \\
          NGC 5205 &  13:30:04 &  62:30:42 &  SBbc &      28.8 &  3.4 &   9.9 &       8.6 \\
          NGC 5406 &  14:00:20 &  38:54:56 &   SBb &      77.3 & 13.1 &  11.0 &       9.7 \\
          NGC 6497 &  17:51:18 &  59:28:15 &  SBab &      87.6 &  8.5 &  11.0 &       9.8 \\
\hline 
\end{tabular}
}
\caption*{Col. (1): Galaxy ID. Col. (2): Right ascension. Col. (3): Declination. Col. (4): Morphological type.  Col. (5): Distance from the NASA-Sloan Atlas Catalog Col. (6): Effective radius from the NASA-Sloan Atlas Catalog, Col. (7): Stellar Mass from the \code{Pipe3D} analysis. Col. (8): Molecular gaseuos mass from the \code{Pipe3D} analysis.}
\end{table*}

\section{Geometric and kinematic galaxy parameters}
\label{Sec:Parameters}

\subsection{Geometric parameters from photometry}
\label{Sec:Geometric}

The TW-method requires a well-constrained orientation of the galaxy disc. To this purpose, it is important to use deep images with high S/N. A nominal mean surface brightness limit for the SDSS images is about 24.5 $\mathrm{mag \, arcsec^{-2}}$ while the typical brightness distribution in the outer regions of galaxies is usually traced below 26 $\mathrm{mag \, arcsec^{-2}}$. At brighter levels, other structural components like spiral arms and rings may still have a non-negligible contribution in the outer regions of galaxies. For the r-band images of the DESI legacy image surveys, a median $5\sigma$ point source depth of 23.4 mag, translates into a median surface brightness limit of $27$  $\mathrm{mag \, arsec^{-2}} $ for a $3\sigma$ detection of an extended source \citep{Schlegel2015}, allowing us to have a more robust probe of the disc PA in the external regions of the selected galaxies. 

\begin{figure*}
	\centering
	\begin{subfigure}{0.33 \textwidth}
	\caption*{\Large \hspace{.5cm} manga-8439-6102}
		\includegraphics[width = \textwidth]{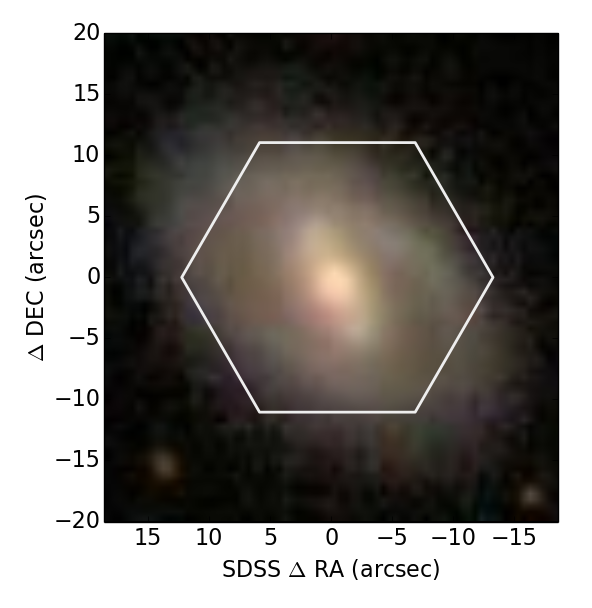}
	\end{subfigure}%
   	\begin{subfigure}{0.33 \textwidth}
   		\caption*{\Large \hspace{.5cm} manga-8135-6103}
		\includegraphics[width = \textwidth]{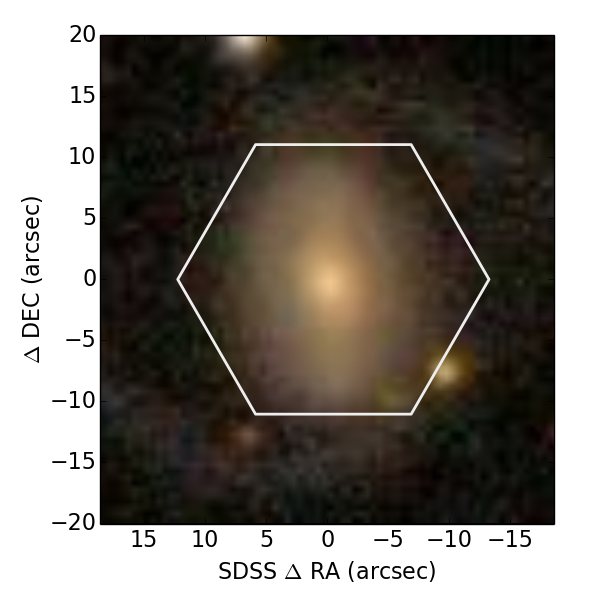}
	\end{subfigure}%
   	\begin{subfigure}{0.33 \textwidth}
   		\caption*{\Large \hspace{.5cm} manga-8341-12704}
		\includegraphics[width = \textwidth]{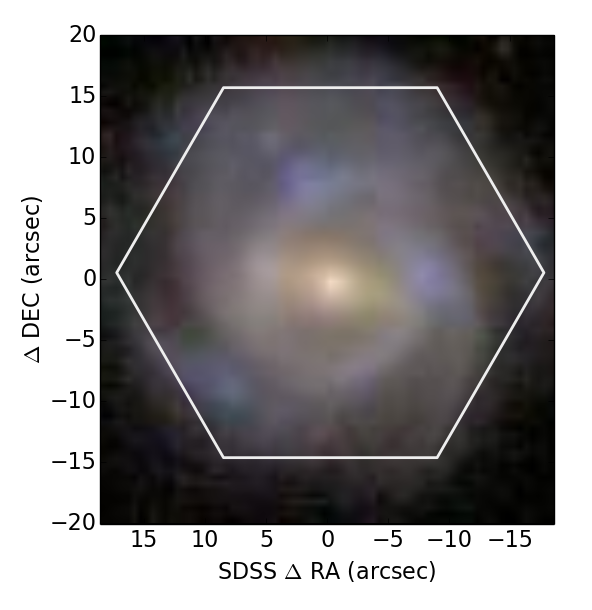}
	\end{subfigure}
    \begin{subfigure}{0.33 \textwidth}
		\includegraphics[width =  \textwidth]{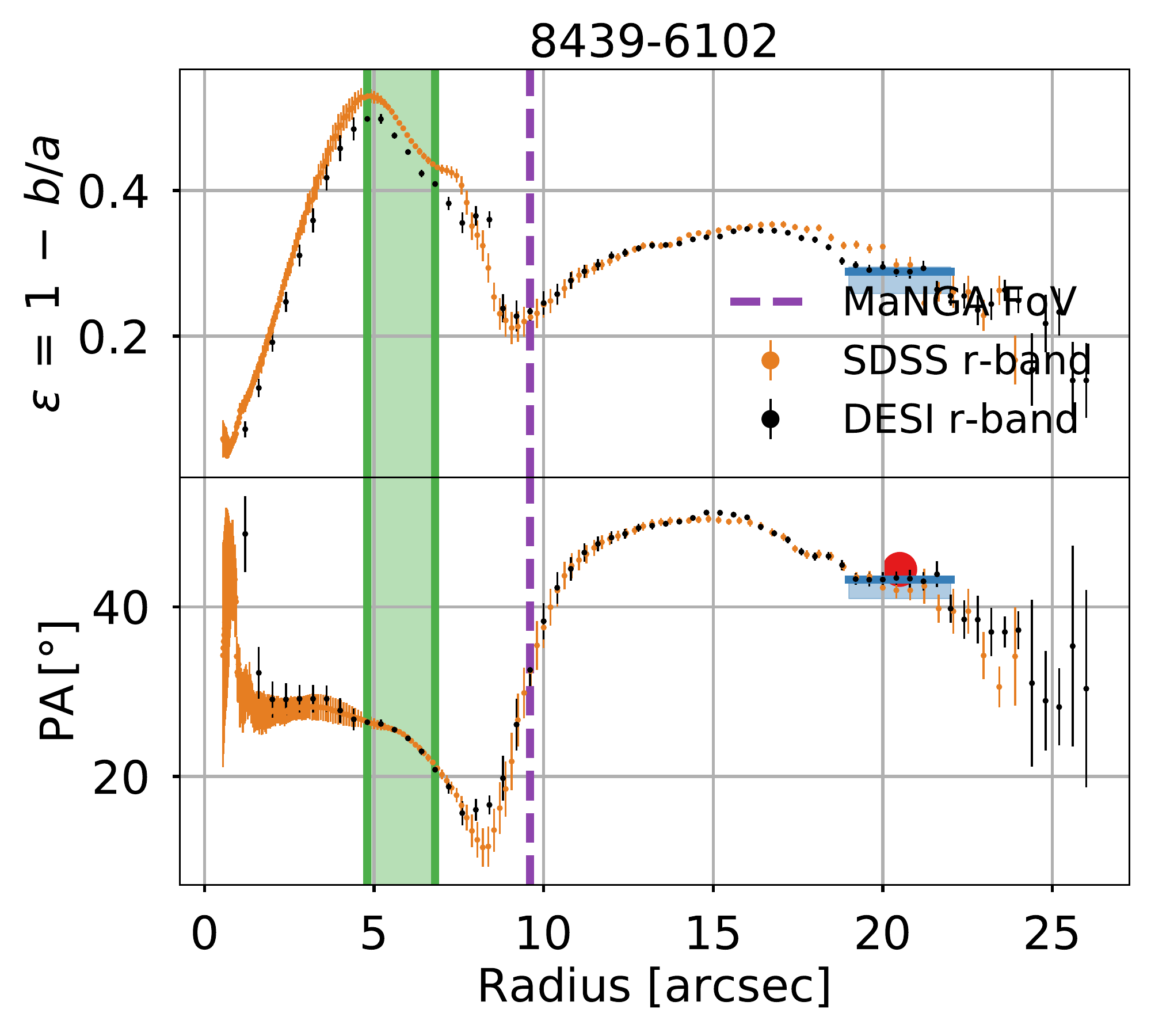}
    \end{subfigure}%
    \begin{subfigure}{0.33 \textwidth}
		\includegraphics[width =  \textwidth]{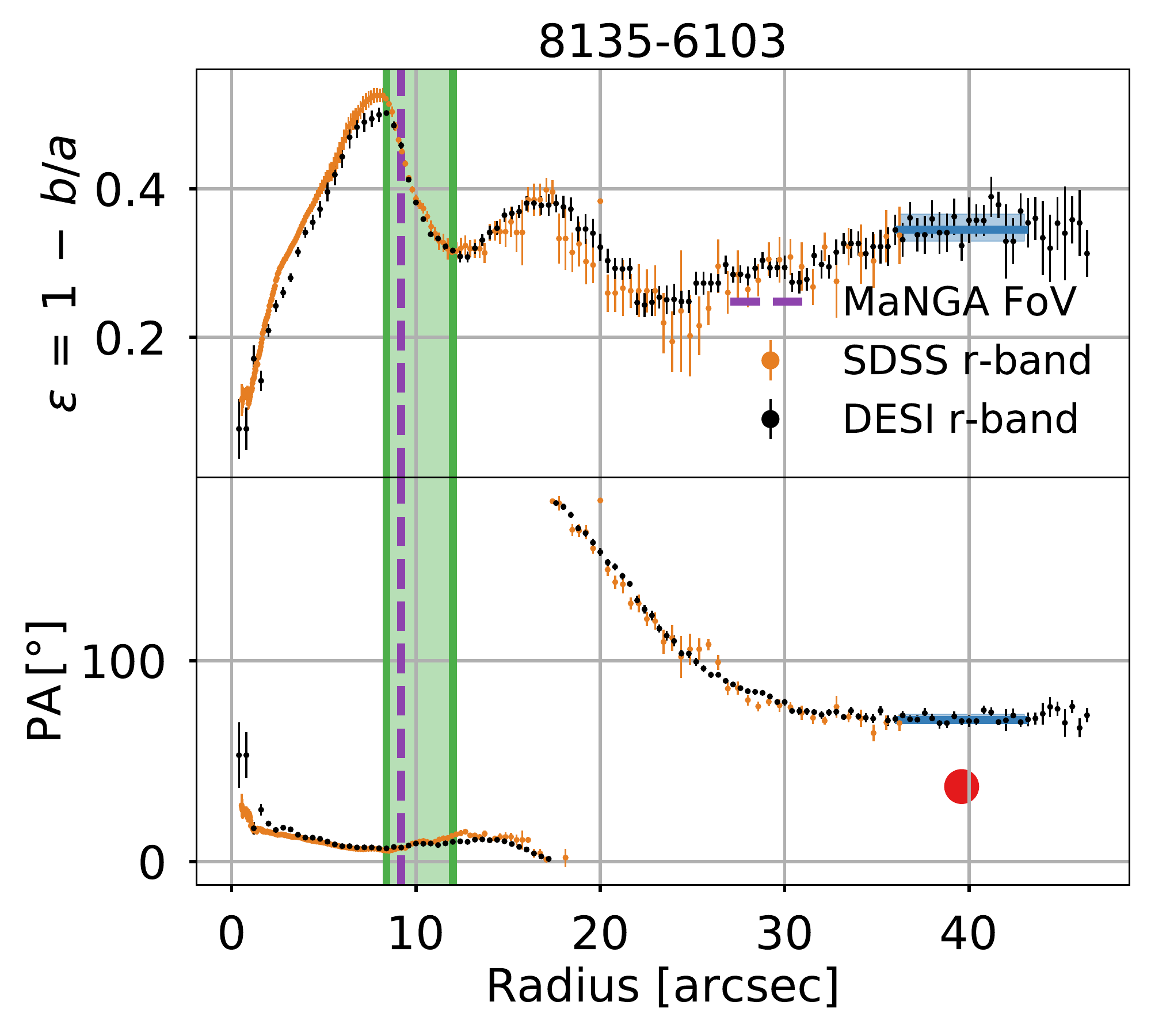}    
    \end{subfigure}%
    \begin{subfigure}{0.33 \textwidth}
		\includegraphics[width =  \textwidth]{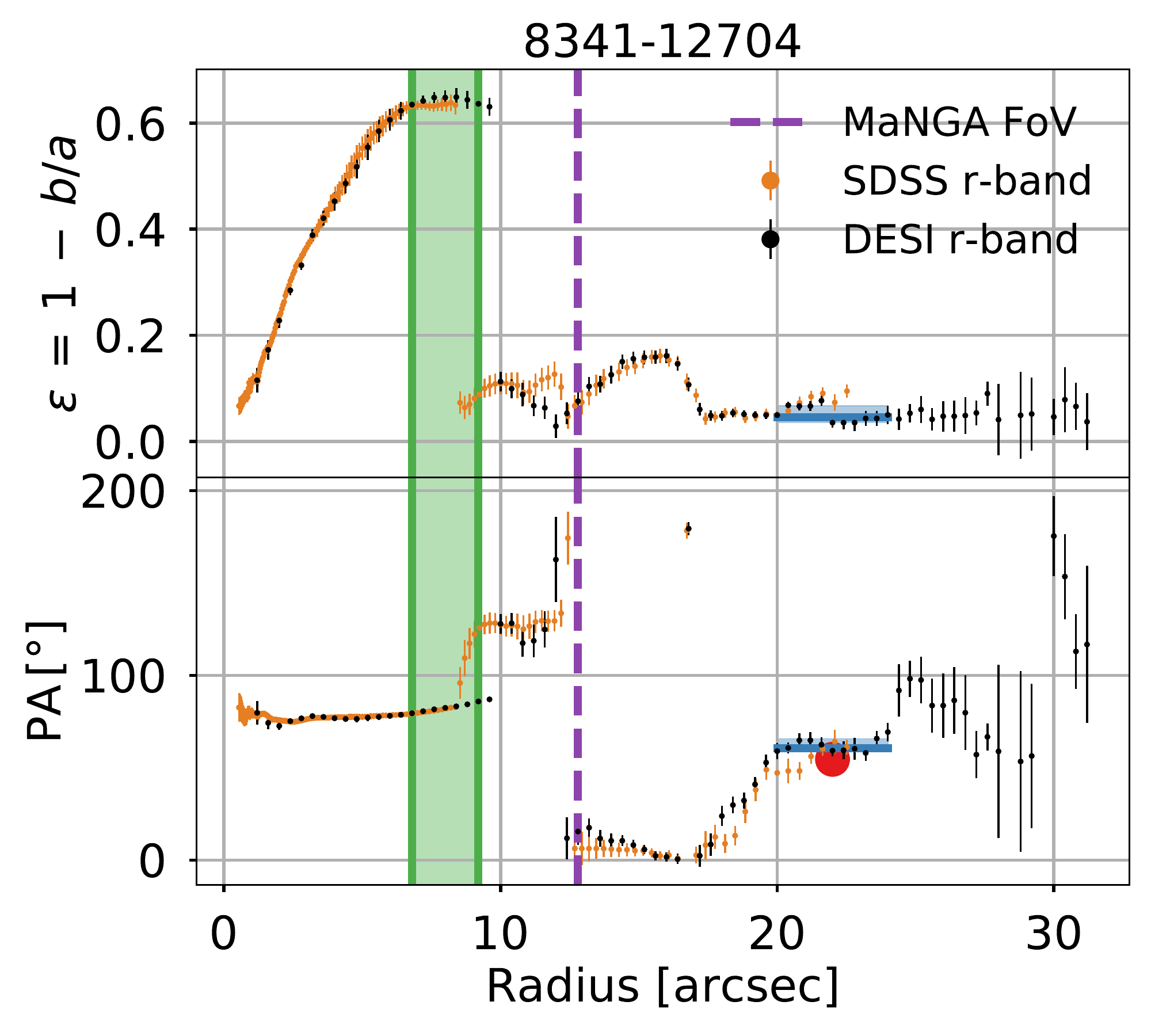}    
    \end{subfigure}
	\caption{\textit{Top panels}: SDSS poststamp of our example galaxies: manga-8439-6102, manga-8135-6103, manga-8341-12704. The white hexagon shows the area observed by the MaNGA IFU. \textit{Bottom panels}: Ellipticity and PA profiles of the legacy DESI and SDSS r-band images, shown in black and orange, respectively. The green region shows the bar radius obtained from the local maximum ellipticity ($R_{bar,1}$) and the change of PA by 5 degrees relative to the PA at $R_{bar,1}$ ($R_{bar,2}$). The purple dashed line shows the approximate FoV of the MaNGA hexagon. The blue horizontal region shows the disc ellipticity and PA (\PAph{}), over the isophotes used for estimating them. The red dot in the PA profiles shows the \PAkn{}. In cases like manga-8439-6102 and manga-8341-12704 where the PA profile flattens at different regions, we choose the one whose PA is closer to \PAkn{}}
   	\label{Fig:Iso_Profiles}
\end{figure*}

We performed an isophote analysis using the ELLIPSE routine \citep{Jedrzejewski1987} within the IRAF environment \citep{Tody1986, Tody1993} to the archive SDSS and DESI legacy r-band images after applying careful masking of field stars (small galaxies in some cases) and other image features in the neighbourhood of the disc. In Figure \ref{Fig:Iso_Profiles}, we compare the ellipticity and PA profiles obtained from both the SDSS and DESI images (in blue and black respectively) in the three example galaxies.Measurements were carried out on the profiles considering a S/N = 3. The approximate length of the MaNGA FoV is shown with a vertical purple segmented  line.

The position of the first maximum in the ellipticity profile is a lower limit to the bar radius \citep[e.g.,][]{Wozniak1995, Michel2006}. We will refer to this radius as $\mathrm{R}_{bar,1}$. 
An upper limit of the bar length is the radius at which the PA profile changes $5 \degree$ for the value at the $\mathrm{R}_{bar,1}$. \citep[e.g.][]{Wozniak1995, Marinova2007, Aguerri2009, Aguerri2015}. We will refer to this radius as $\mathrm{R}_{bar,2}$. In Figure \ref{Fig:Iso_Profiles} these two measurements delimit the green shaded region. We deproject the bar length using the analytic method described by \cite{Gadotti2007}. The method assumes the bar can be described as a simple ellipse and has uncertainties $\sim 10 \%$ at moderate inclination angles $(i \leq 60 \degree)$ \citep{Zou2014}. We will refer to the corresponding deprojected values as $\mathrm{R}^{dep}_{bar,1}$ and $\mathrm{R}^{dep}_{bar,2}$.

The measurement of the disc orientation is not always straightforward. The presence of field stars, disc warps, outer rings, companion galaxies and outer spiral arms can alter such estimate. In practice, the choice of a representative outer disc region is somewhat arbitrary, and the estimation of \PAph{} can change from author to author. In those situations, a correct estimation of the orientation parameters should be consistent with those coming from kinematic data. However, there are numerous examples in the literature where galaxy orientation parameters coming from photometric and kinematic data are different. Several examples have been found using IFS data; \citep[e.g.][]{Emsellem2004, Krajnovic2011, Barrera2014, Barrera2015, Allen2015a, Jin2016}. Our approximation to this problem is to further constrain a-posteriori the \PAph{} by using as a prior knowledge of the \PAkn{} (which is discussed in the next section). This is illustrated in Figure \ref{Fig:Iso_Profiles} in the PA profile of our example galaxies manga-8439-6102 and manga-8341-12704. In both cases, there are two regions where the PA profile flattens (around 15 and 20 arcsec in both cases). We choose the region whose PA is closer to the \PAkn{} which is denoted with a red dot. Our final measurement of \PAph{} and inclination is the median and 1-sigma percentiles weighted by the intrinsic error given by \code{ELLIPSE} enclosed by the chosen region. This results in asymmetrical errors which are shown in Table \ref{Tab:Geometric_parameters}.

Even though the legacy images of DESI are deeper than the corresponding to SDSS, the difference between \PAph{} is negligible in most of our sample. Only in two galaxies (manga-7495-12704 and manga-8257-3703), there is a noticeable difference of $\sim 5 \degree$ between both estimations of the \PAph{}.

\subsection{Rotation curve derivation}
\label{Sec:Rotation_curve}

Analyzing the kinematics of barred galaxies has several complications including bar induced non-circular streaming, projection effects and random motions \citep{Valenzuela2007}. These perturbations produce a twist in the inner disc velocity field and complicate the estimation of the rotation curve (hereafter $V(R)$), and \PAkn{}. To minimize such problems, we derived the rotation curve using the \code{Velfit} software applied directly to the $\mathrm{H}_\alpha$ velocity maps produced by \code{Pipe3D}.

The code \code{Velfit}, developed by \cite{Spekkens2007} and  \cite{Sellwood2010}, fits the best combination of circular streaming and bi-symmetric non-circular motions pixel by pixel in the velocity field, correcting for projection effects. Assuming a flat disc and a distortion with a fixed orientation, \code{Velfit} tries to fit the velocity map expressed in a Fourier series using a Levenberg-Marquardt algorithm as a minimization technique. This routine minimizes the $\chi^2$ maximum likelihood estimator by using a gradient search through parameter space. However, least-squares approaches are highly sensitive to the first approximation guesses of the fittings. When applied to under-sampled data sets, in which many local minima in the $\chi^2$ space are present, the routine could get easily trapped in those local minima. In this work, we used a modified version of \code{Velfit}, where the Markov Chain Monte Carlo technique is implemented. That provides a method of surveying the parameter space that rapidly converges to the posterior probability distribution of the input parameters using the Metropolis-Hastings algorithm (Aquino-Ortíz in prep.). 

We applied such methodology to galaxies in our sample except for manga-7990-12704 which is an early type galaxy with no significant $\mathrm{H}_\alpha$ emission. Instead, for manga-7990-12704 we applied \code{Velfit} to the stellar velocity field. The procedure corrects for projection effects and non-circular streaming, however, in contrast with gaseous components where random motions are damped because of cooling, stellar kinematics requires correction because of random motions. Such correction is not obvious for barred galaxies where common corrections underestimate such effect \citep{Valenzuela2007}. Because building a detailed dynamical model is beyond this paper scope, we warn that dynamical interpretation of such galaxy should be considered with caution.

\begin{figure*}
	\centering
    \begin{subfigure}{0.33 \textwidth}
    \includegraphics[width= 1.05\textwidth]{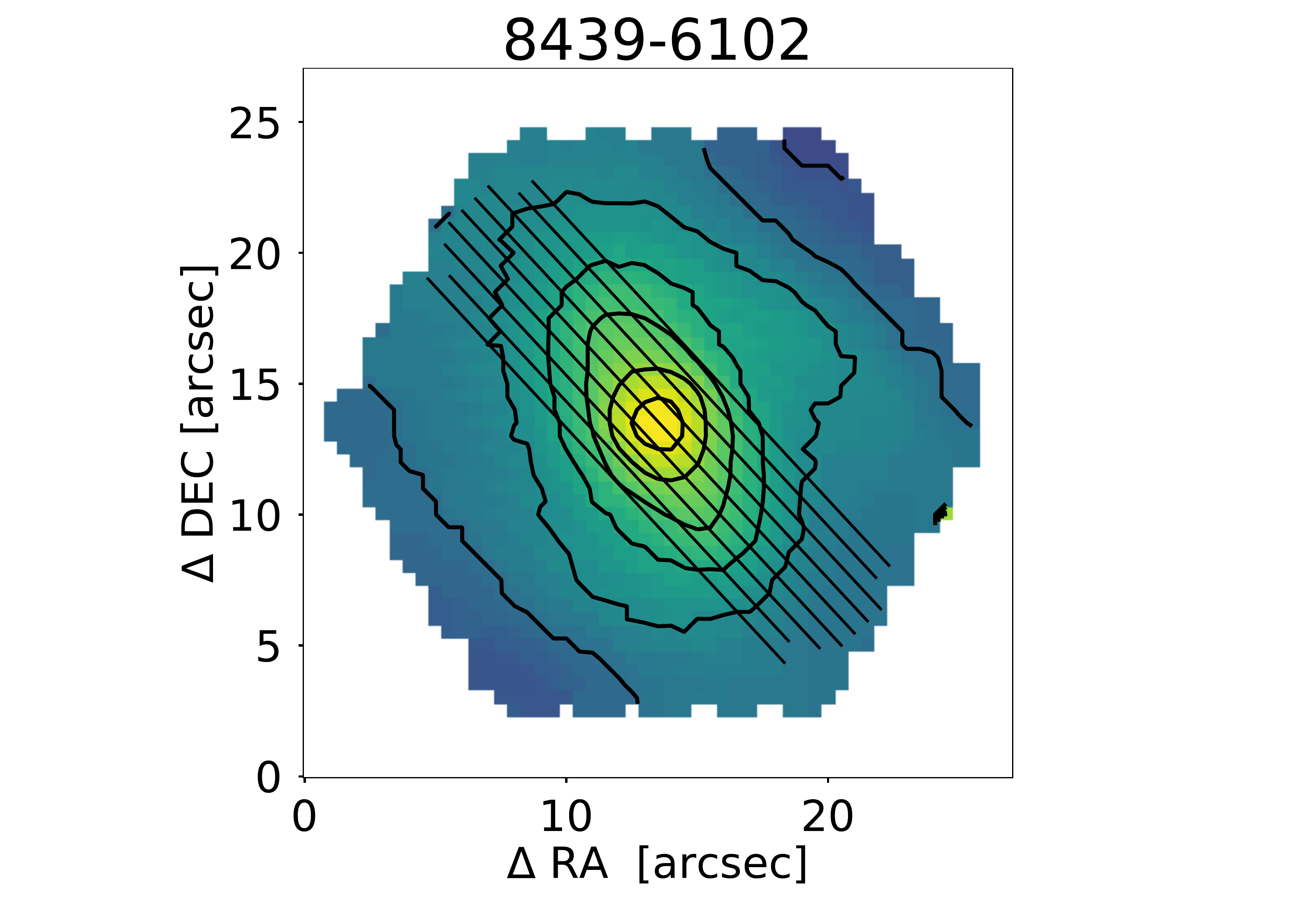}
    \end{subfigure}%
    \begin{subfigure}{0.33 \textwidth}
	\includegraphics[width= 1.05\textwidth]{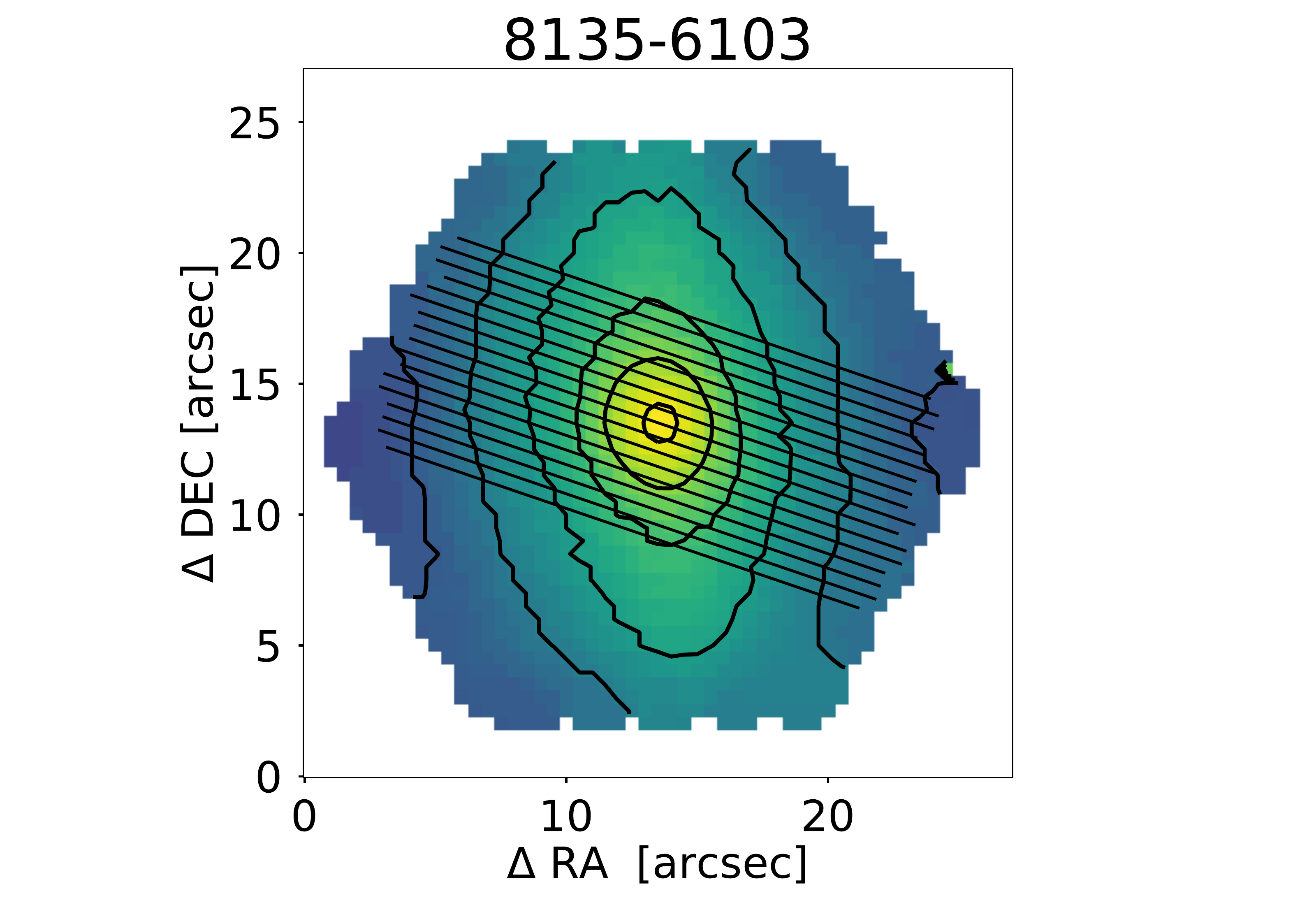}
    \end{subfigure}%
    \begin{subfigure}{0.33 \textwidth}
   	\includegraphics[width= 1.05\textwidth]{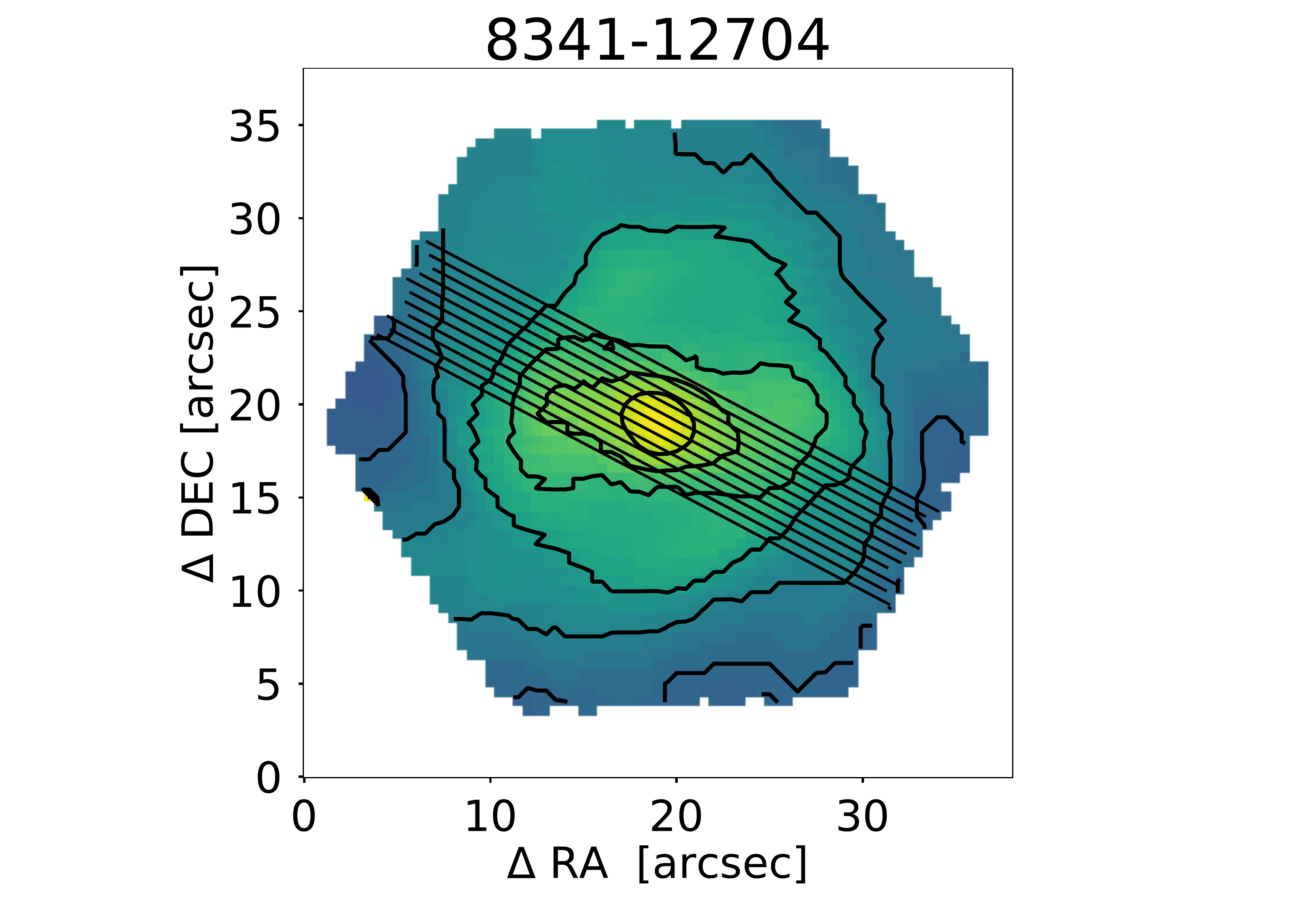}
    \end{subfigure}
	\begin{subfigure}{0.33 \textwidth}
		\includegraphics[width = 1.05\textwidth]{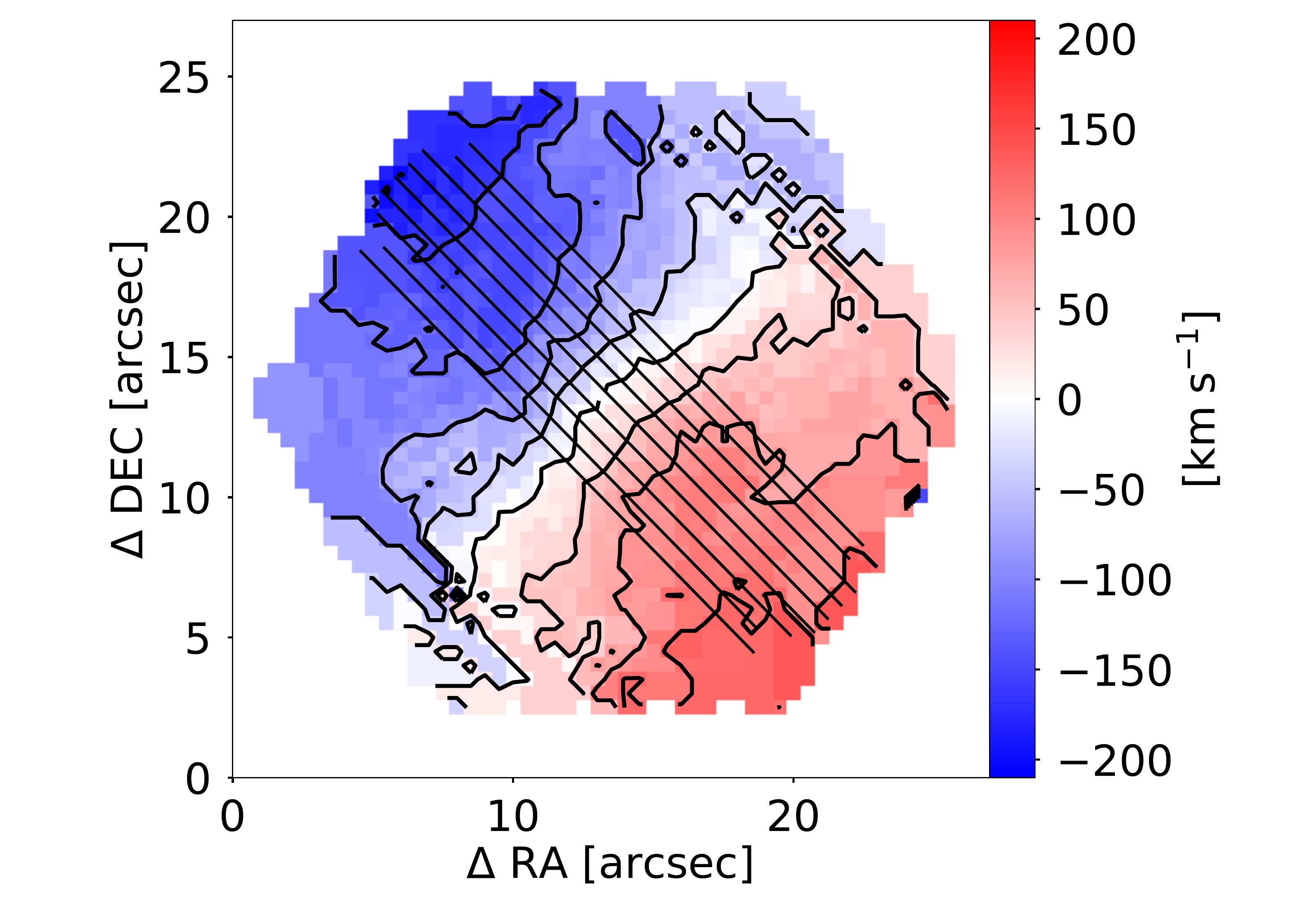}
	\end{subfigure}%
   	\begin{subfigure}{0.33 \textwidth}
		\includegraphics[width = 1.05\textwidth]{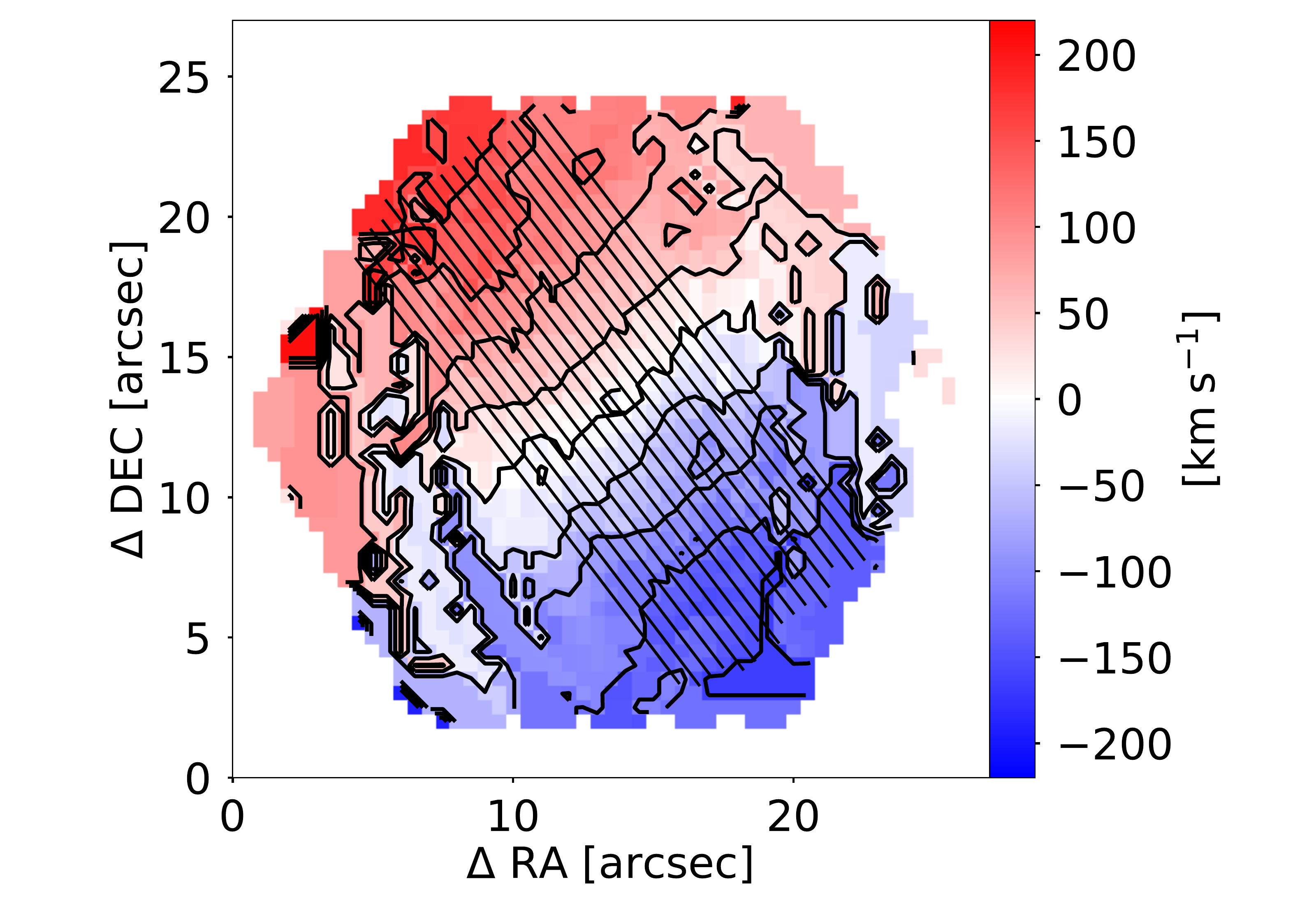}
	\end{subfigure}%
   	\begin{subfigure}{0.33 \textwidth}
		\includegraphics[width = 1.05\textwidth]{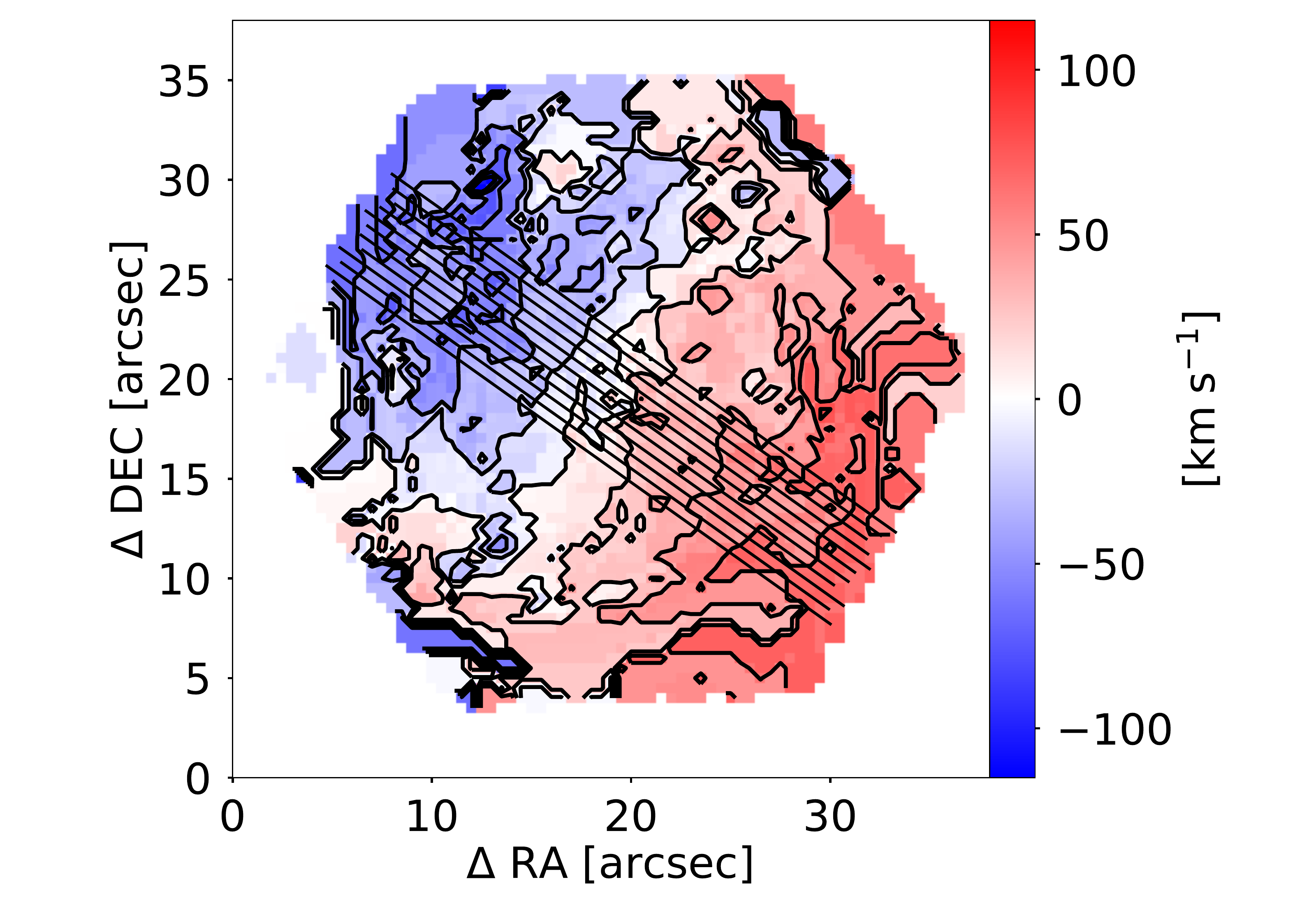}
	\end{subfigure}
    \begin{subfigure}{0.33 \textwidth}
		\includegraphics[width =  0.91 \textwidth]{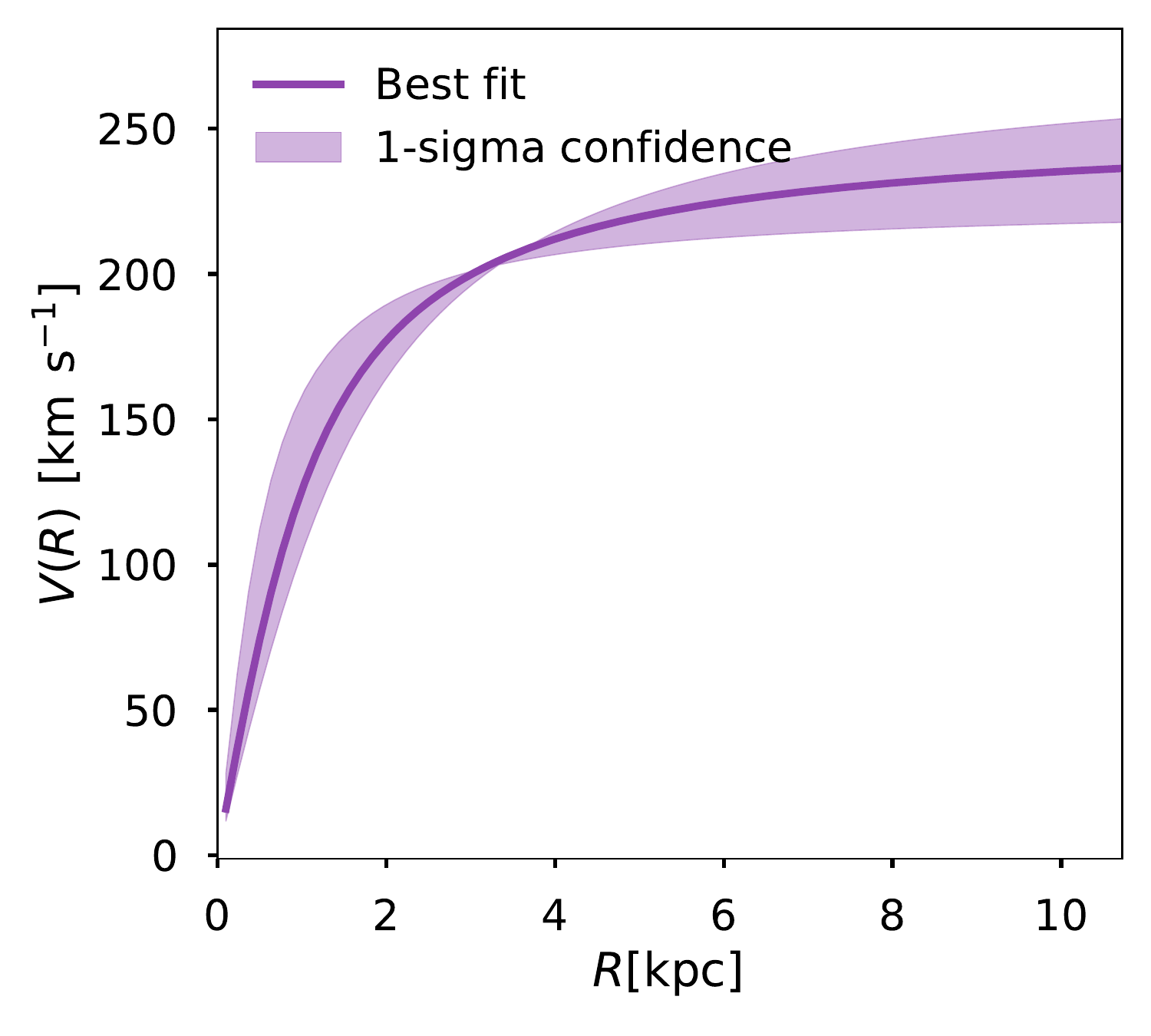} 
    \end{subfigure}%
    \begin{subfigure}{0.33 \textwidth}
		\includegraphics[width =  0.91 \textwidth]{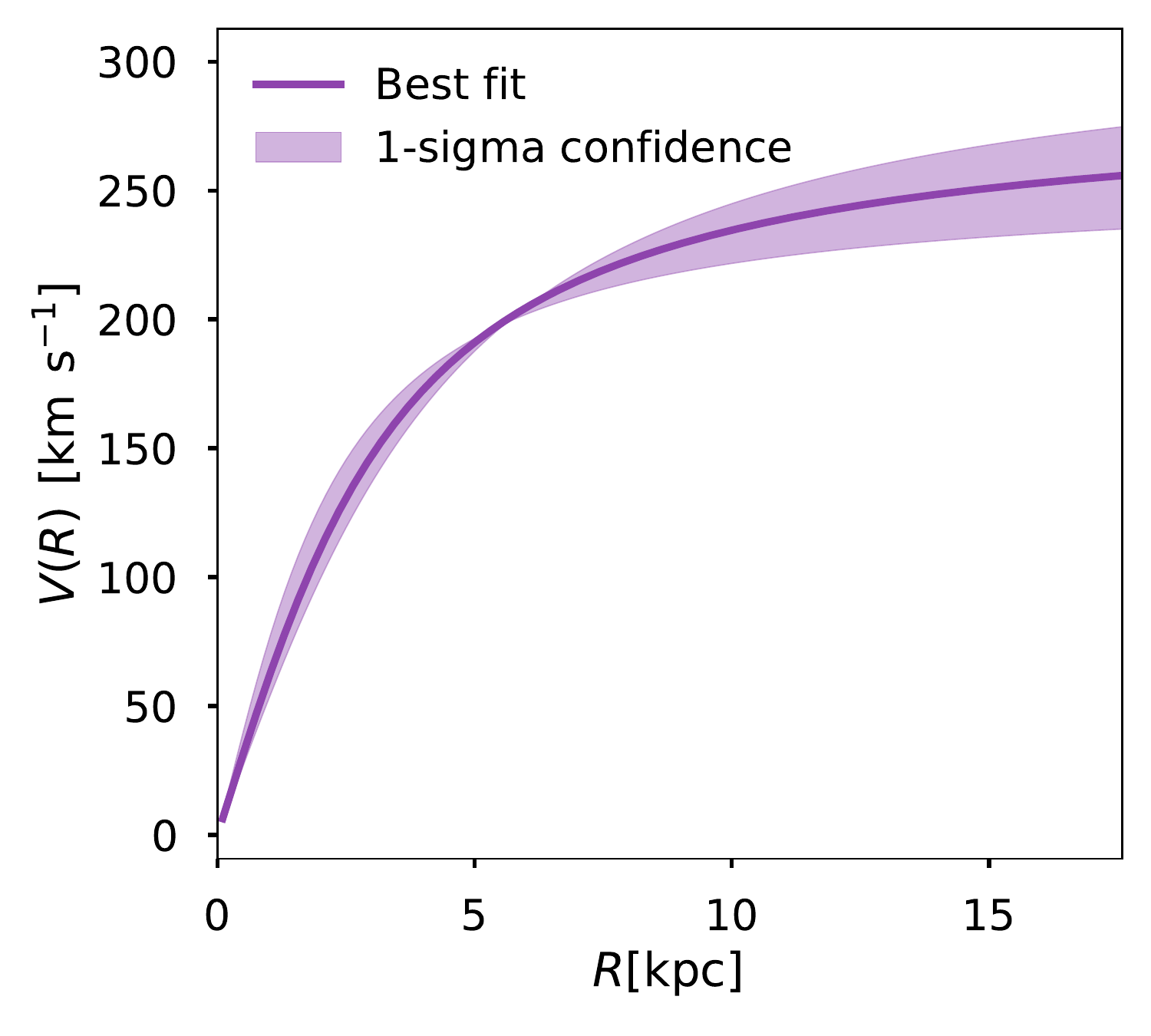} 
    \end{subfigure}%
    \begin{subfigure}{0.33 \textwidth}
		\includegraphics[width =  0.91 \textwidth]{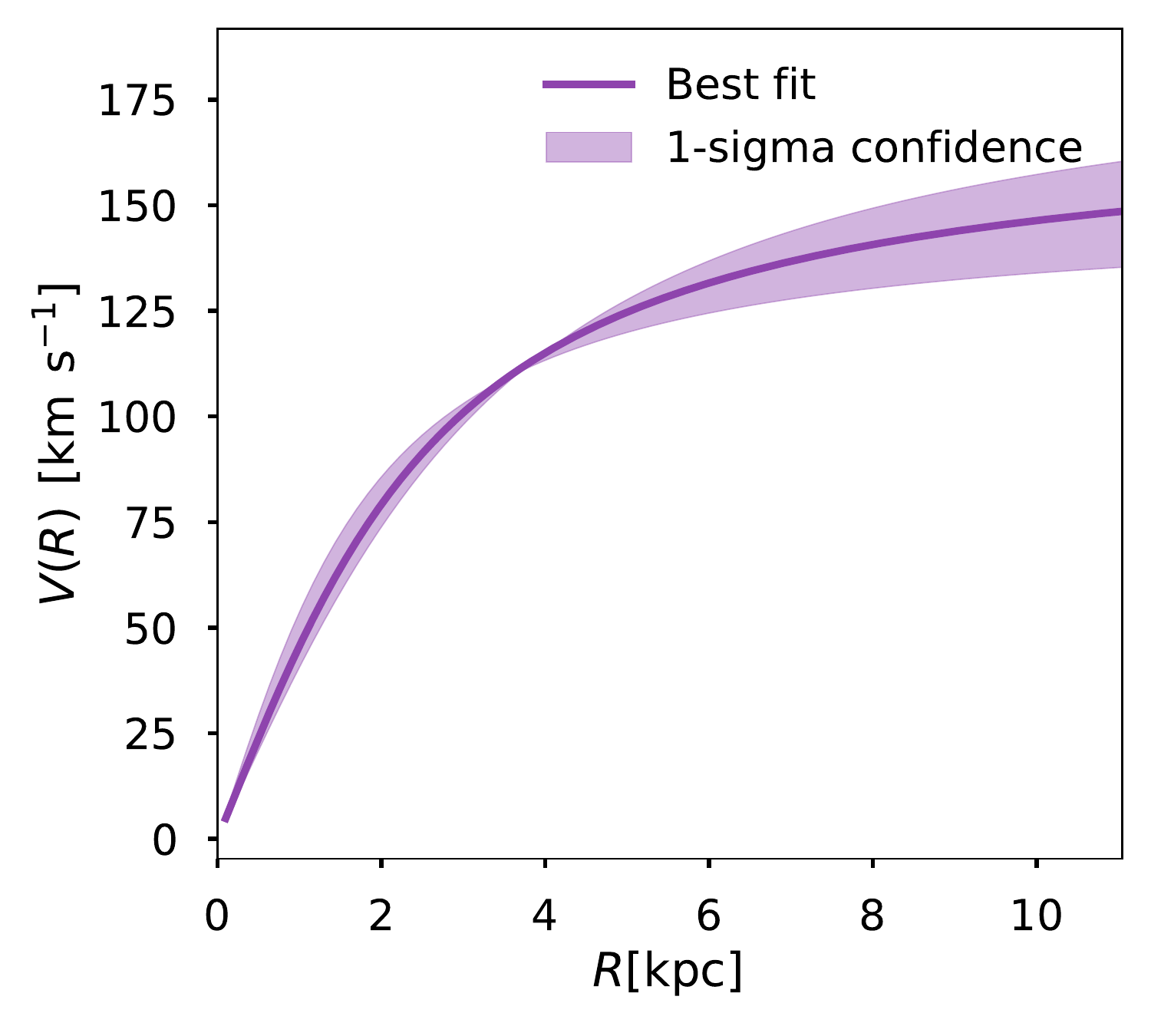}    
    \end{subfigure}   
	\caption{ \textit{Top panels}: Stellar flux maps derived by \code{Pipe3D} for our example galaxies. The black straight lines show the pseudo-slits oriented towards \PAph{}.
    \textit{Middle panels}: Stellar velocity maps derived by \code{Pipe3D}. Slits are oriented towards \PAkn{}. The difference between PA's is specially notorious in manga-8135-6103. 
	\textit{Bottom panels}: Rotation curves modeled by \code{Velfit} using the H$_\alpha$ velocity maps derived from \code{Pipe3D}. The shaded region shows the 1-sigma confidence interval.}
   	\label{Fig:Rot_curves}
\end{figure*}

To extrapolate the \code{Velfit} velocities to further radii we modeled $V(R)$ with a basic 2-parameters function shown in equation \ref{Eq:arctan}  \citep[Equation 1 of ][]{Courteau1997}. This simple model assumes that the rotation curve flattens at a transition radius $r_t$, and reaches an asymptomatic velocity $V_{flat}$. In Figure \ref{Fig:Rot_curves} we show the fitted rotation curves of our example galaxies. In Table \ref{Tab:Geometric_parameters} we show the fitted parameters $V_{flat}$ and $r_t$.

\begin{equation}
V(r)=V_{sys} + \frac{2}{\pi} \: V_{flat} \: \arctan \left( \frac{r-r_0}{r_t} \right)
\label{Eq:arctan}
\end{equation}

\begin{table*}
\centering
\caption{Geometric and kinematic parameters of our sample}
\label{Tab:Geometric_parameters}
{
\renewcommand{\arraystretch}{1.3}
\begin{tabular}{ccccccccccc}
\hline \hline
 Galaxy & $i$ & \PAph &\PAkn &  $\text{PA}_{bar}$ &$\mathrm{R}_{bar,1}$ &  $\mathrm{R}_{bar,2}$ &  $\mathrm{R}^{dep}_{bar,1}$ & $\mathrm{R}^{dep}_{bar,2}$ & $V_{flat}$ & $r_t$ \\
     &  [\degree] & [\degree] & [\degree] &[\degree]  & [arcsec]   & [arcsec]   &  [kpc]   & [kpc] & [km s$^{-1}$ ] & [kpc] \\
 (1) & (2) & (3) & (4) & (5) & (6) & (7) & (8) & (9) & (10) & (11) \\  
 \hline
 manga-7495-12704 &  $50.8^{+1.9}_{-1.4}$ &  $173.0^{+1.4}_{-1.5}$  &  $172.5 \pm 1.3$ &    143.0 &   5.2 &   8.8 &       3.7 &       5.9 &   $248.8 \pm 3.7$ &  $1.7 \pm 0.1$ \\
 manga-7962-12703 &  $63.8^{+2.0}_{-1.6}$ &   $36.6^{+1.0}_{-2.2}$  &   $33.4 \pm 1.0$ &     50.4 &  12.0 &  14.0 &      13.5 &      16.1 &  $350.0 \pm 29.0$ &  $3.0 \pm 1.3$ \\
 manga-7990-12704 &  $41.7^{+1.3}_{-5.7}$ &   $74.9^{+4.2}_{-2.8}$  &   $68.1 \pm 1.4$ &     81.5 &  11.2 &  12.8 &       6.2 &       7.1 &  $273.3 \pm 39.0$ &  $4.1 \pm 1.1$ \\
  manga-8135-6103 &  $49.1^{+1.6}_{-1.2}$ &   $70.5^{+2.9}_{-1.1}$  &   $37.3 \pm 2.5$ &      6.5 &   8.4 &  12.0 &       9.8 &      14.5 &  $284.9 \pm 31.9$ &  $2.8 \pm 0.9$ \\
 manga-8243-12704 &  $51.2^{+0.3}_{-0.8}$ &   $23.3^{+1.1}_{-0.5}$  &   $23.0 \pm 0.8$ &     29.3 &   5.2 &   7.6 &       2.5 &       3.7 &   $222.4 \pm 9.0$ &  $2.9 \pm 0.3$ \\
  manga-8256-6101 &  $46.6^{+5.7}_{-5.4}$ &  $123.8^{+3.0}_{-5.0}$  &  $135.9 \pm 2.7$ &     59.3 &   6.0 &   7.2 &       4.1 &       4.9 &  $297.5 \pm 45.4$ &  $1.0 \pm 0.8$ \\
  manga-8257-3703 &  $56.2^{+1.4}_{-0.3}$ &  $152.6^{+0.5}_{-1.1}$  &  $157.6 \pm 1.3$ &    132.7 &   5.2 &   7.2 &       3.2 &       4.9 &  $236.8 \pm 16.3$ &  $1.0 \pm 0.3$ \\
 manga-8312-12704 &  $42.0^{+1.7}_{-2.7}$ &   $28.0^{+3.1}_{-2.9}$  &   $34.8 \pm 4.8$ &    150.8 &   7.2 &   9.6 &       5.4 &       7.2 &   $185.0 \pm 5.1$ &  $1.7 \pm 0.2$ \\
  manga-8313-9101 &  $39.1^{+0.1}_{-0.2}$ &  $108.3^{+1.1}_{-3.0}$  &  $111.7 \pm 2.0$ &    154.7 &   3.2 &   4.0 &       2.9 &       3.8 &   $283.2 \pm 9.2$ &  $1.0 \pm 0.3$ \\
 manga-8317-12704 &  $67.7^{+1.1}_{-2.1}$ &  $106.9^{+0.5}_{-0.4}$  &  $111.3 \pm 1.8$ &    126.3 &   6.4 &   8.0 &      12.8 &      15.9 &   $299.6 \pm 8.5$ &  $1.0 \pm 0.4$ \\
 manga-8318-12703 &  $54.7^{+1.2}_{-1.9}$ &   $53.7^{+0.6}_{-1.2}$  &   $57.6 \pm 0.8$ &     86.5 &   5.2 &   6.4 &       5.5 &       7.4 &   $284.6 \pm 5.5$ &  $1.0 \pm 0.1$ \\
 manga-8341-12704 &  $17.3^{+3.7}_{-2.1}$ &   $60.7^{+5.1}_{-1.6}$  &   $54.6 \pm 4.5$ &     79.5 &   6.8 &   9.2 &       4.2 &       5.6 &  $170.2 \pm 21.5$ &  $2.2 \pm 0.7$ \\
  manga-8439-6102 &  $44.7^{+0.5}_{-2.5}$ &   $43.2^{+0.4}_{-2.3}$  &   $44.4 \pm 0.5$ &     26.4 &   4.8 &   6.8 &       3.4 &       5.1 &  $251.2 \pm 26.8$ &  $1.0 \pm 0.5$ \\
 manga-8439-12702 &  $54.8^{+0.7}_{-0.8}$ &   $30.6^{+0.9}_{-0.7}$  &   $36.1 \pm 2.6$ &    145.7 &   6.0 &   7.2 &       5.5 &       6.4 &  $273.9 \pm 16.5$ &  $1.0 \pm 0.3$ \\
 manga-8453-12701 &  $36.9^{+1.7}_{-2.7}$ &   $98.5^{+0.6}_{-0.2}$  &  $104.7 \pm 0.8$ &     43.3 &   5.6 &   6.4 &       3.3 &       3.7 &   $187.3 \pm 8.5$ &  $2.2 \pm 0.3$ \\
          NGC5205 &  $55.3^{+0.8}_{-0.5}$ &  $170.7^{+0.6}_{-0.7}$  &  $167.5 \pm 0.5$ &    109.6 &  11.2 &  12.8 &       2.6 &       2.9 &  $204.2 \pm 14.0$ &  $1.3 \pm 0.2$ \\
          NGC5406 &  $46.4^{+0.5}_{-0.7}$ &  $110.8^{+1.8}_{-1.9}$  &  $119.7 \pm 1.4$ &     56.5 &  18.0 &  20.8 &       8.9 &      11.2 &  $283.8 \pm 16.1$ &  $1.0 \pm 0.5$ \\
          NGC6497 &  $62.0^{+0.7}_{-1.0}$ &  $111.1^{+0.6}_{-1.8}$  &  $110.9 \pm 5.7$ &    152.6 &   8.8 &  12.8 &       6.6 &       9.6 &  $296.3 \pm 25.9$ &  $1.4 \pm 0.9$ \\
\hline 
\end{tabular}
}
\caption*{Col. (1): Galaxy. Col. (2): Disc inclination Col. (3): Disc photometric PA. Col. (4): Disc kinematic PA. Col. (5): Bar PA. Col. (6): Maximum ellipticity bar radius. Col. (7): $\Delta PA = 5 \degree$ bar radius. Col. (8): Deprojected inner bar radius. Col. (9): Deprojected outer bar radius.  Col. (10) Flatten circular velocity. Col. (11) Transition radius}
\end{table*}

\subsection{Number, separation and length of the pseudo slits}
\label{Sec:Number_Sep_Leng}

Choosing the optimal number of slits for the application of the TW-method is not trivial. Increasing the number of slits leads to a better fitting of \Om{} in the \Vint{} vs. \Xint{} diagram. However, as more slits are placed at outer radii, the effect of the bar is diluted in comparison to other components of the galaxy. Doing so may result in values that interrupt the linear trend between \Vint{} and \Xint{}. As a compromise of these arguments the number of pseudo slits we use per galaxy correspond to the minimum between the following two numbers (i) the maximum number of slits that can be fitted inside the inner bar radius $R_{bar,1}$, and (ii) the maximum number of slits that preserve the linear trend between \Vint{} and \Xint{}.

The separation between slits should be wide enough to prevent the same pixel to be summed by two consecutive slits. If the slits are aligned, either horizontally or vertically to the 2D array of pixels, a separation of 1 pixel is enough to prevent summing the same information. However, in the most likely scenario in which the slits have a different orientation the separation should be increased by a factor $\cos ^{-1} \theta$ where:

\begin{equation}
\theta =
   \begin{cases}
    PA,      & \text{if} \;\;\;\; PA < 45 \degree \\
    PA-90,   & \text{if} \;\;\;\; 45 \degree < PA < 135 \degree \\
    PA-180,  & \text{if} \;\;\;\: PA > 135 \degree
   \end{cases}
\end{equation}

Formally the TW-integrals are required to be symmetrical and as large as possible since the integration limits go from $-\inf$ to $\inf$. However, since the data from MaNGA and CALIFA galaxies are embedded in hexagons, the maximum slit length depends on the orientation and centre of each slit. This is especially important when the observed galaxy is not centred. Based on the previous arguments we let each slit to extend from their centre up to the closer edge of the hexagon in the direction of their orientation. This preserves the symmetry of the integrals while keeping the slits as long as they can be.We will refer to this slit length as $L_{max}$. In Section \ref{Sec:Pseudo-slits-length} we discuss the error associated with the slit length.

\section{Measuring the error from different sources}
\label{sec:Errors}

\subsection{The centring error}
\label{Sec:centring_error}

In principle, the symmetry of the surface brightness and the velocity maps makes the TW-method insensitive to the centring errors. However, if the galaxy disc is rich in substructure, or presents an excess of luminosity in the central regions, the centre choice could affect the measurement of \Om{}. We estimate the centre of our galaxies by performing a Monte Carlo (MC) simulation of barycenter of the stellar flux maps, using the random flux errors estimated by \code{Pipe3D} and a random radius around the brightest pixel. The resulting centroid uncertainty in our sample is $\sim 0.5$ pixels on average. 

For this test, we repeated the TW-integrals times by randomly changing the centre of the galaxy using a 2D-gaussian distribution with a standard deviation of $\sigma = 1 \, \text{pixel}$, which is slightly higher than the mean centroid accuracy. We then measure the median and 1-sigma percentiles of \Om{}. We will refer to the 1-sigma percentiles obtained from this test as \dCp{} and \dCm{} and the average error as \dC{}.

Table \ref{Tab:Omega} shows the resulting centring error for each galaxy. On average, this error accounts for $ \sim 5 \%$ of the relative error, which, as expected is almost negligible. Nonetheless,  this is the dominant error in two galaxies (manga-8257-12704 and NGC 5205) and is greater than 10 \% in two more (manga-8243-12704 and manga-8341-12704). These last two, present a clear excess of luminosity in the central region.

\subsection{The PA error}
\label{Sec:PA_error}

On average, the associated error to the disc PA is $1.7 \degree$ and $2.0 \degree$ for \PAph{} and \PAkn{} respectively. These errors are intrinsic to the methodology, however, they could be biased by outer non-axisymmetric structures. Our example galaxy, manga-8135-6103 illustrates such systematic, with strong spiral arms that extend to the outermost region, biasing the \PAph{} (see Figure \ref{Fig:Iso_Profiles}). This results in a difference with \PAkn{} of $33.2 \degree$. The frequency of multiple non-axisymmetric structures in the outer regions of disc galaxies and their impact on a correct estimate of a disc orientation deserves a more detailed study to properly take into account such systematic. 

Three galaxies present problems with the kinematic orientation obtained from the best \code{Velfit} model (manga-8341-12704, manga-7990-12704 and NGC 5406). The example galaxy manga-8341-12704 has two major complications. (1) The low inclination angle of $17.3 \degree$, which makes the kinematic information more uncertain; (2) The resolved kinematics traced by $H_\alpha$ show a perturbation near the centre of the galaxy. The origin of this perturbation may be related to the presence of a strong AGN (Cortés-Suárez et al. in prep), which cannot be modelled by \code{Velfit}. The galaxy manga-7990-12704, as we mentioned before, has a poor quality $H_\alpha$ emission, so we used the stellar velocity map instead, which, is affected by random stellar motions. In the case of NGC 5406, the spiral arms that start at the bar region could have a strong non-axisymmetric contribution, that could bias the \code{Velfit} model. 

To estimate the associated error we repeated the measurement of \Om{} for 500 different angles, equally spaced between $\min ($\PAph, \PAkn$) - 10 \degree$ and $\max($\PAph, \PAkn$) + 10 \degree$. This interval allows to study the behavior of \Om{} inside the uncertainties of our PA measurements.

\begin{figure*}
	\centering
    \begin{subfigure}{0.33 \textwidth}
		\includegraphics[width =  \textwidth]{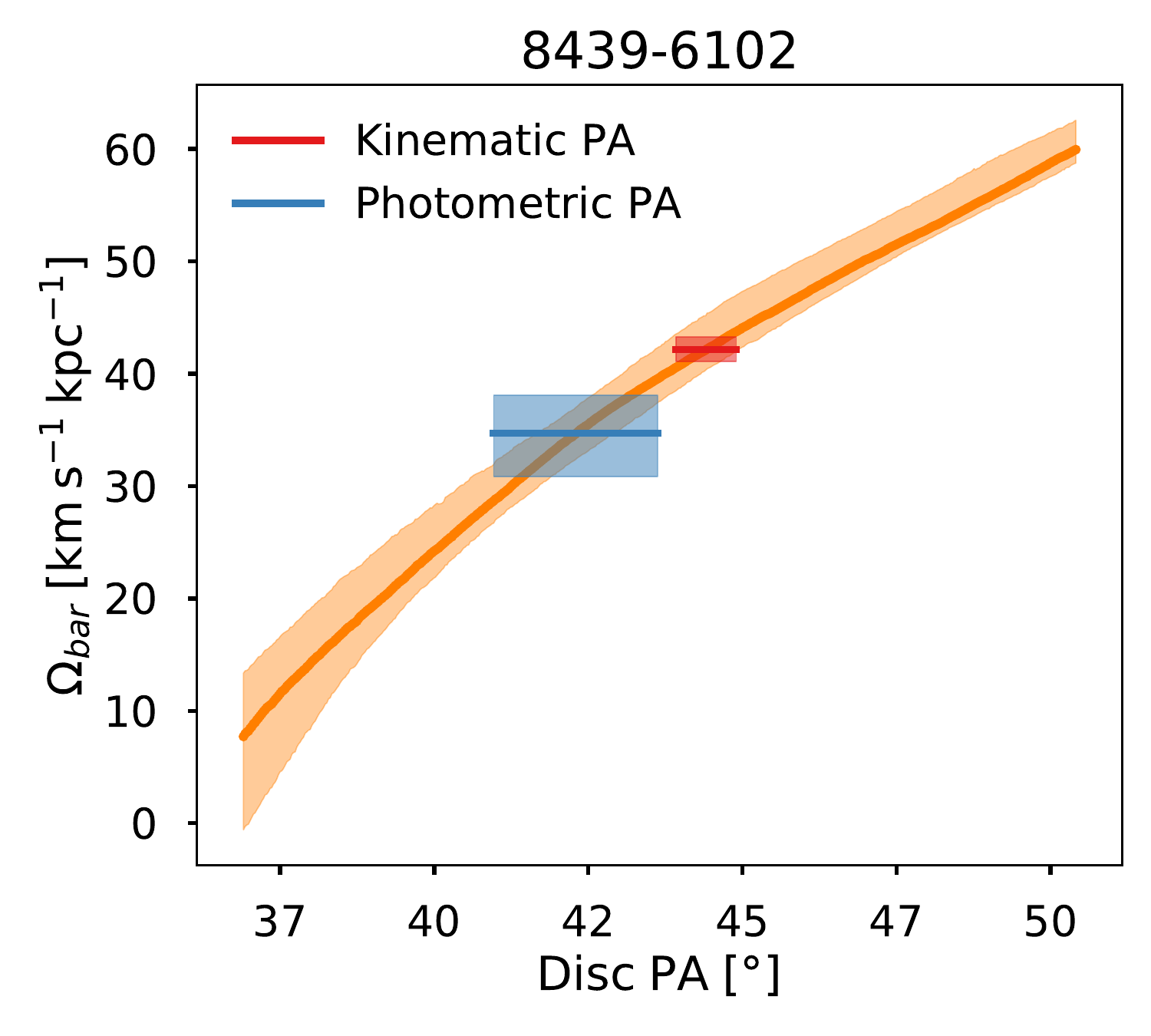}
    \end{subfigure}%
    \begin{subfigure}{0.33 \textwidth}               
		\includegraphics[width =  \textwidth]{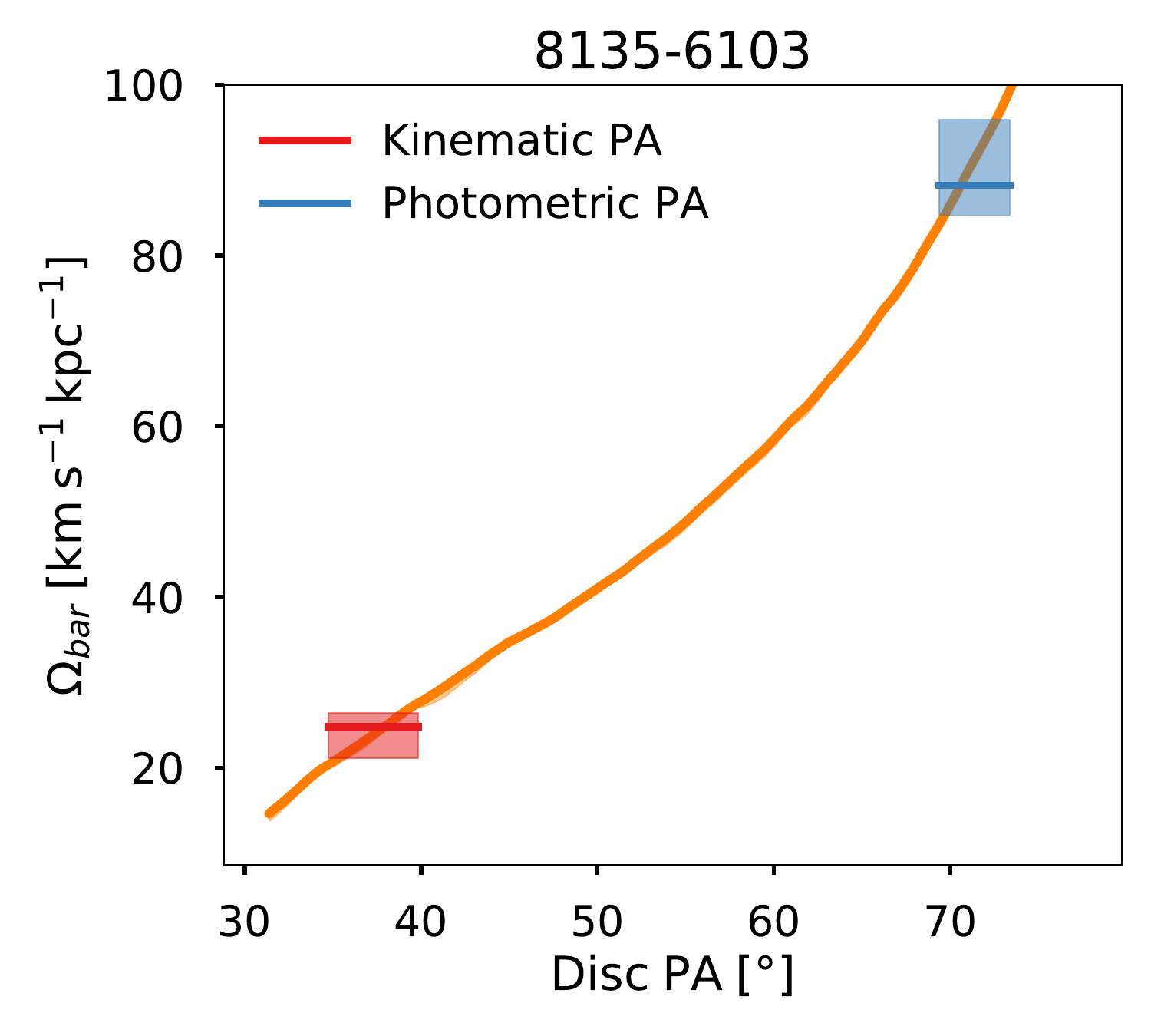}
    \end{subfigure}%
    \begin{subfigure}{0.33 \textwidth}
		\includegraphics[width =  \textwidth]{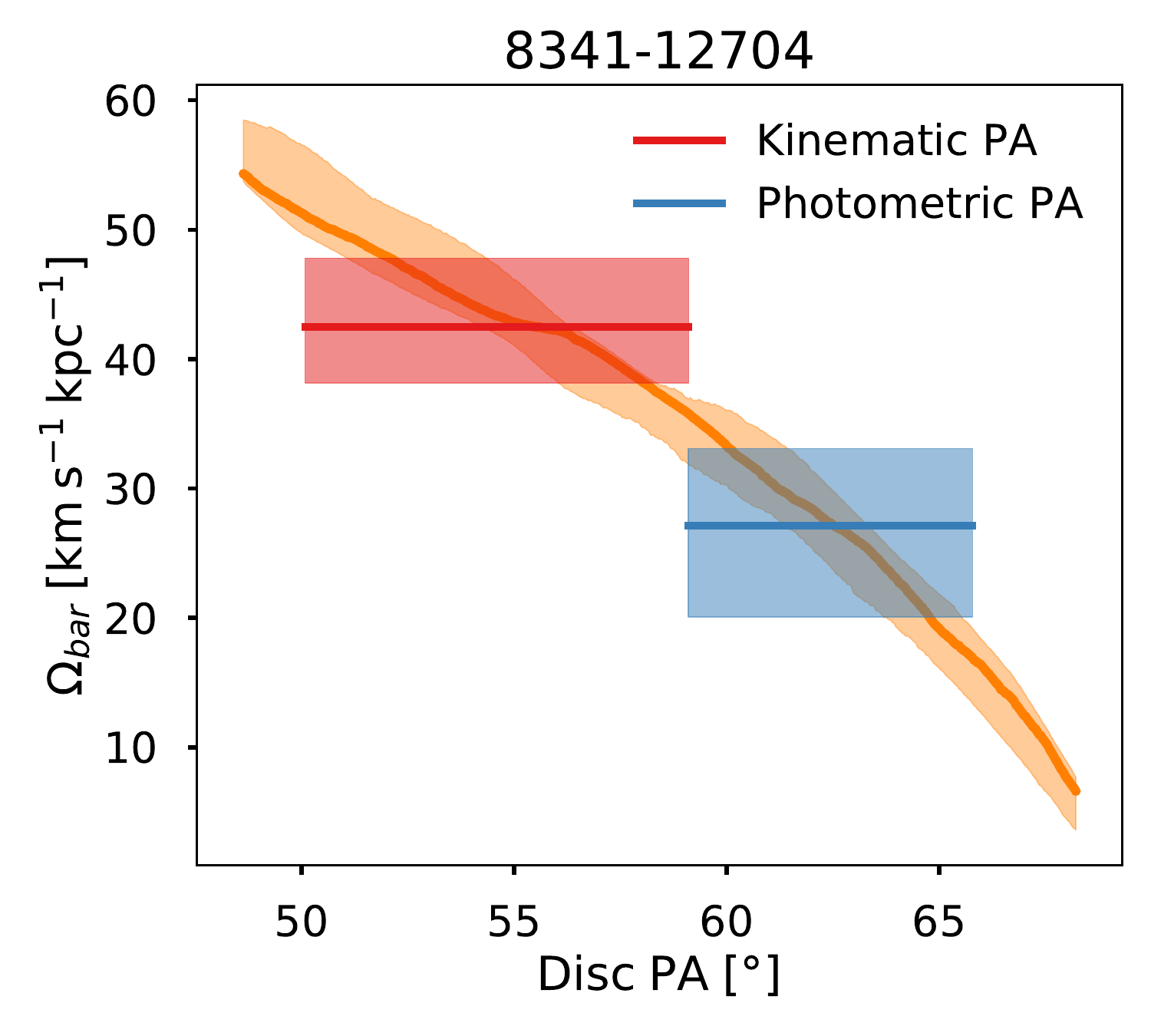} 
     \end{subfigure}
  \caption{Bar pattern speed vs. PA of our example galaxies. At each PA we repeat the measurement of \Om{} 1000 times changing slightly the centre as described in Section \ref{Sec:centring_error} The orange shaded region shows 1-sigma percentiles \dCp{} and \dCm{}. The blue and red region show the measurement of \Omph{} and \Omkn{} enclosed by their 1-sigma percentiles weighted by the \dC{} respectively. In the case of manga-8439-6102 \Omph{} and \Omkn{} are similar.} manga-8135-6103 illustrates an example where the photometric PA is heavily affected by its strong spiral arms, making \Omph{} unreliable. manga-8341-12704 is a low inclination galaxy with strong non-axisymmetric motions. The resulting pattern speed \Omkn{} barely crosses the angular velocity curve (see Figure \ref{Fig:Ang_curves}).
   \label{Fig:PA-error}
\end{figure*}

At each angle, we first estimated the centring error with the procedure described in Section \ref{Sec:centring_error}. Figure \ref{Fig:PA-error} shows the \Om{} vs PA plot for our example galaxies. The orange dots show the median value of \Om{} after changing randomly the centre 1000 times. The orange shaded region shows the centring error given by \dCp{} and \dCm{}. One common characteristic we observed in the \Om{} vs PA plots is an asymptotic behaviour of the curves. This behaviour is observed when the orientations of the slits are close to the major or minor axis of the bar. This causes $\langle X \rangle$ to become symmetric and tend to zero. An example can be seen in manga-8135-6103 in Figure \ref{Fig:PA-error}, as the PA of the slits gets closer to the bar minor axis, which lies at $\sim 96 \degree$. 

We will refer as \Omph{} and \Omkn{} to the median value of \Om{} weighted by \dC{} at the regions enclosed by \PAph{} and \PAkn{}. The 1-sigma percentiles will be denoted as \dPAp{} and \dPAm{} and the average error as \dPA{}. This is the dominant error source in 13 galaxies of our sample and accounts on average for $\approx 15 \%$ of the relative error in \Om{}. 

In some galaxies, the estimation of \Omph{} and \Omkn{} provides with non-physical results as they never cross the disk angular velocity curve at any radius, being impossible to recover \CR{}. This is the case of our two example galaxies manga-8135-6103 and manga-8341-12704 with \Omph{} and \Omkn{} respectively (see Figure \ref{Fig:Ang_curves}). In general, \Omph{} proved to be more physically meaningful than \Omkn{}. The value of \Omph{} crosses the angular velocity curve in all our galaxies except for manga-8135-6103, while \Omkn{} fails to do so in 4 galaxies. To simplify the discussion in further sections, from hereafter, we will refer to our photometric results, with the exception of manga-8135-6103.

\subsection{The slits length error }
\label{Sec:Pseudo-slits-length}

Setting the length of the slits to the maximum can be problematic in some scenarios. As we discuss in Figure \ref{Fig:TW_example}, if we do not properly orient the slits with the disc, a fictitious pattern speed will appear. Increasing the length of the slits could amplify that error. Also, the effect of other non-axisymmetric structures such as the spiral arms, on the TW-integrals poorly understood. This is especially important if they rotate with different pattern speeds or occupy a large section of the FoV compared to the bar.

Since MaNGA has a radial coverage of 1.5 and 2.5 $R_e$ ($\sim$ 70\% and $\sim$ 30\% of the sample galaxies respectively), there are cases where the bar covers entirely the observed hexagon (for example manga-8135-6103).  It is not clear if using slits that not cover the entire bar may introduce a systematic error in our measurements.

\begin{figure*}
	\centering
    \begin{subfigure}{0.33 \textwidth}
		\includegraphics[width =  \textwidth]{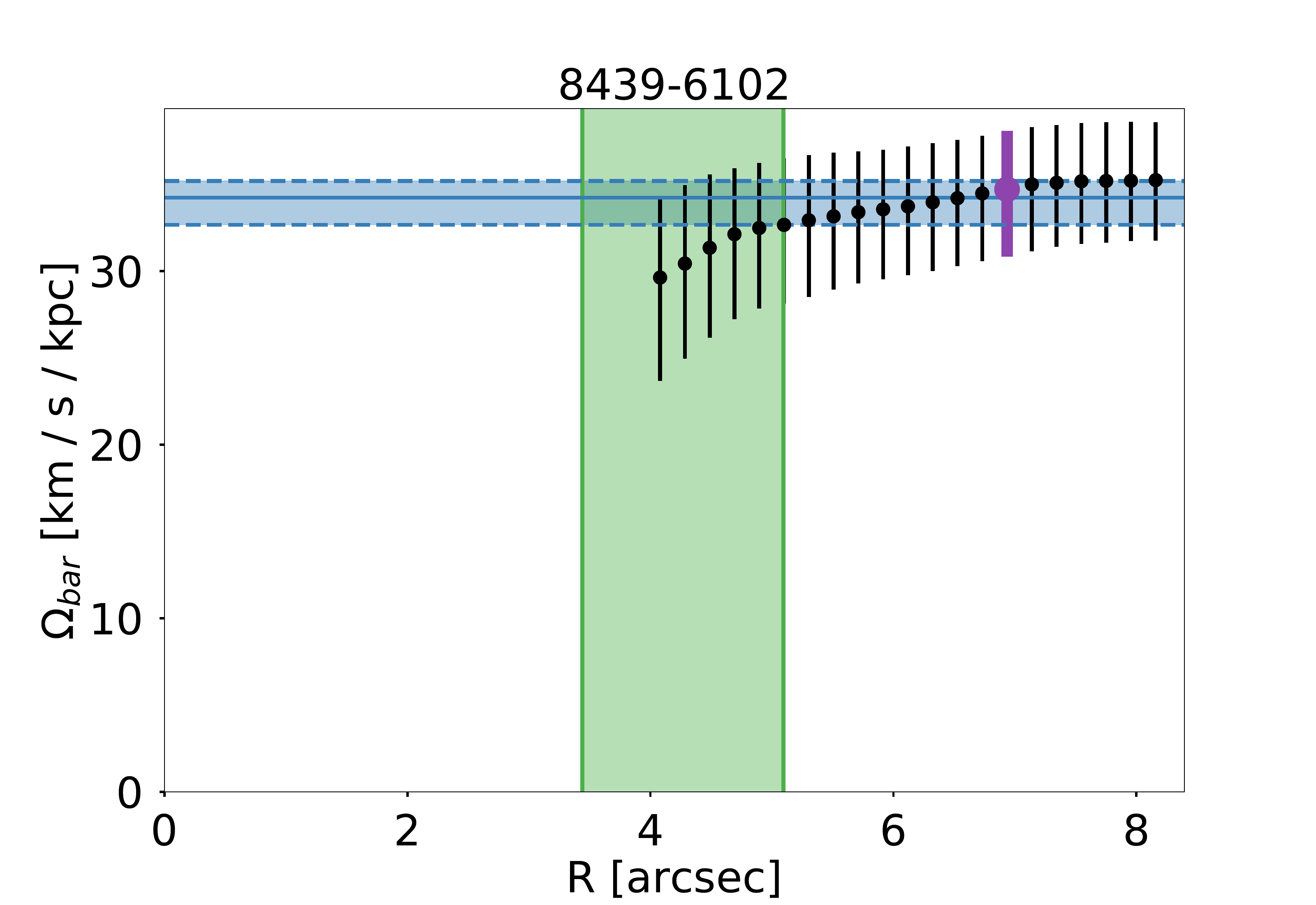}
    \end{subfigure}%
    \begin{subfigure}{0.33 \textwidth}               
		\includegraphics[width =  \textwidth]{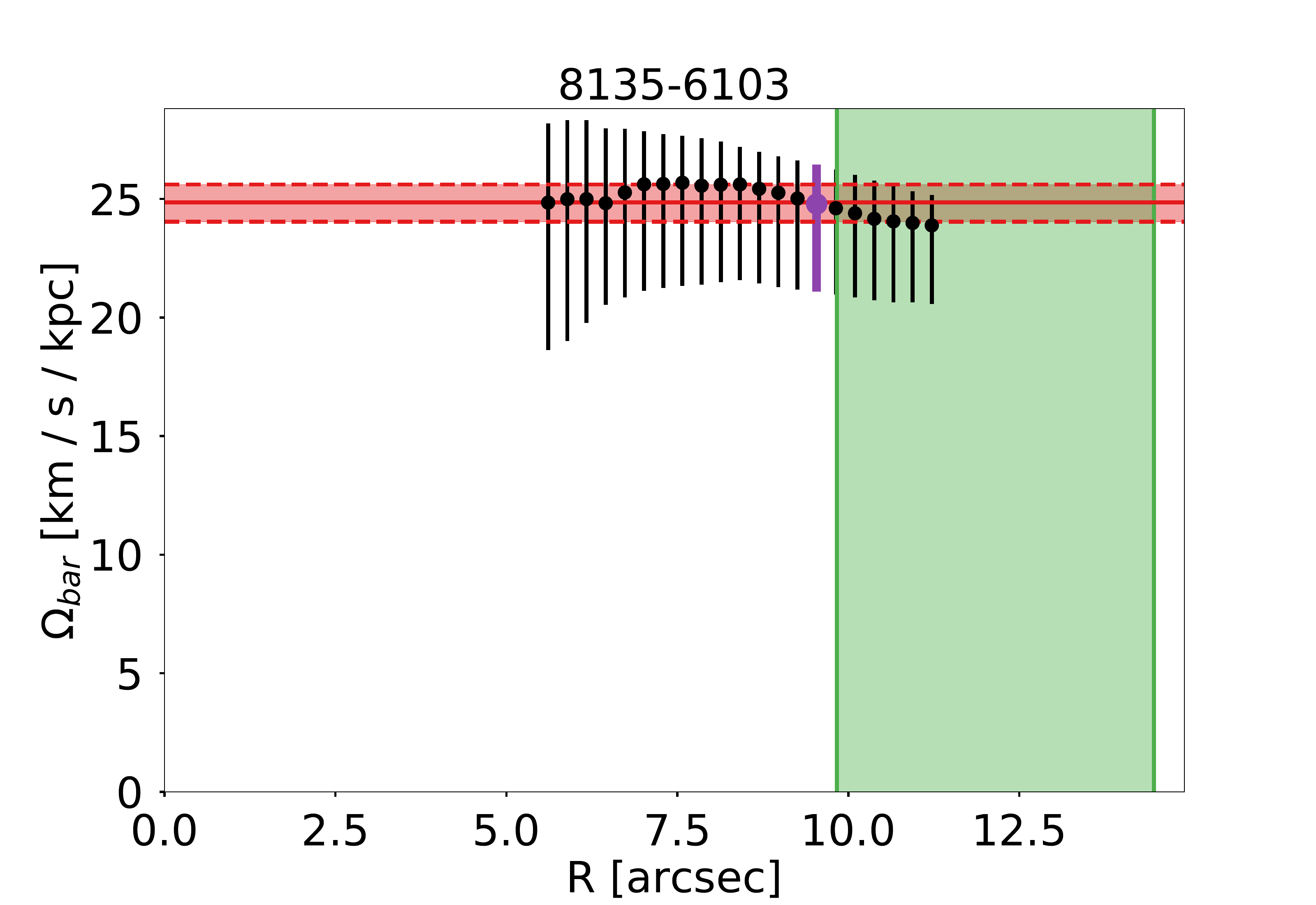}
    \end{subfigure}%
    \begin{subfigure}{0.33 \textwidth}
		\includegraphics[width =  \textwidth]{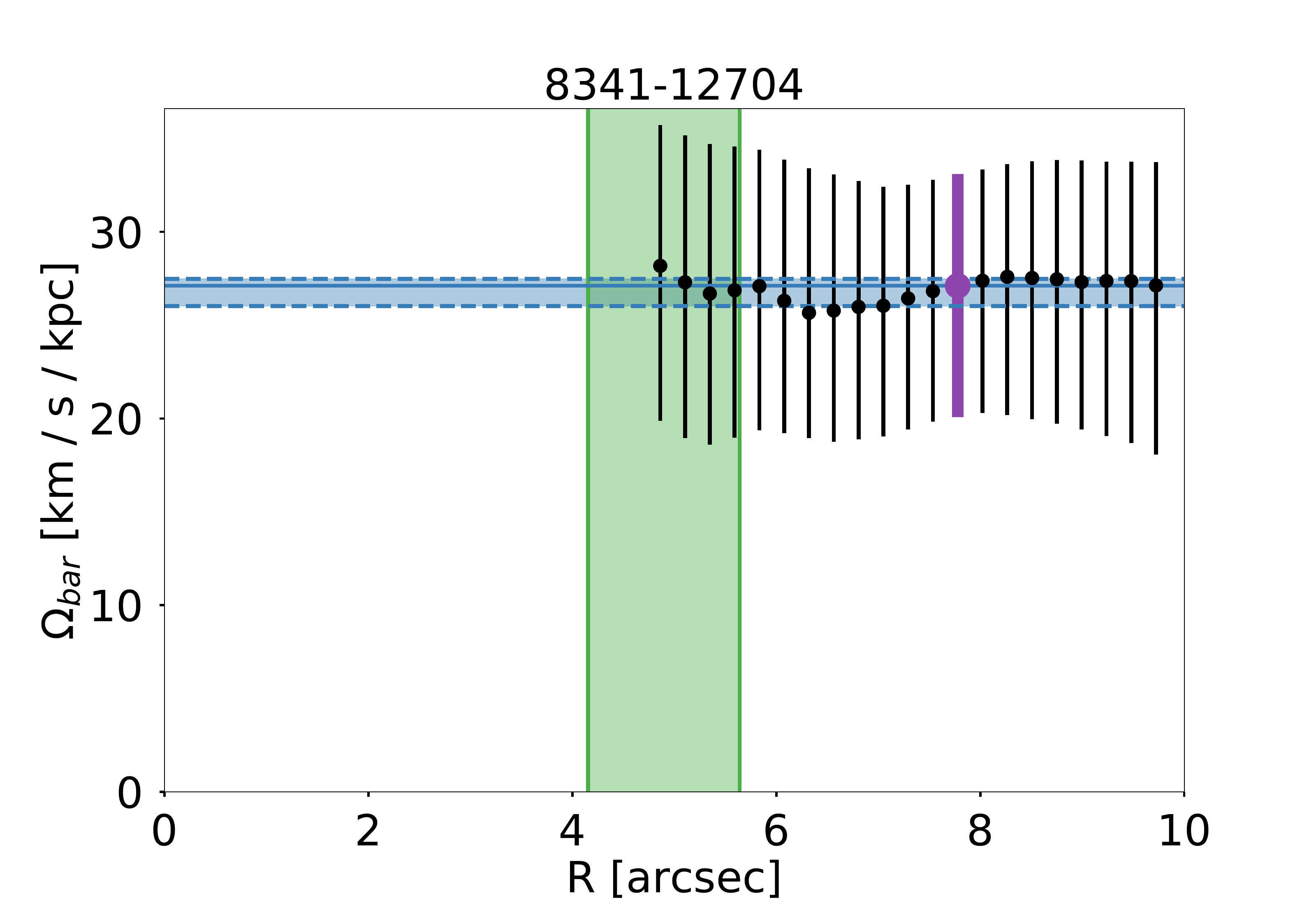} 
     \end{subfigure}
\caption{\Om{} vs. slits length at different radius. The black dots show the measurements of \Om{} at different slits lengths, between $0.5 \; L_{max}$ and $L{max}$. At each slits length we re-estimate the centring and PA errors as explained in Sections \ref{Sec:centring_error} and \ref{Sec:PA_error}. The blue and red regions show the median and 1-sigma error of the weighted \Omph{} and \Omkn{} respectively. Among the 4 measurements closest to the median, we highlight in purple the one with the smallest centring plus PA error. The green region shows the bar radius, as they where shown in Figure \ref{Fig:Iso_Profiles}. }
\label{Fig:Omega_Slits}
\end{figure*}

 As we described in Section \ref{Sec:Number_Sep_Leng}, we let $L_{max}$ be the maximum length each slit can have while preserving the symmetry around its centre. To estimate the slits length error we measure \Omph{} and \Omkn{} for 21 slits lengths, equally spaced, from slits with length $L_{max}$ to $0.5 \;  L_{max}$. In Figure \ref{Fig:Omega_Slits} we show the resulting values of \Omph{} for manga-8439-6102 and manga-8341-12704 and \Omkn{} for manga-8135-6103, repeating the same procedure we described in the previous section. For simplicity, we add in quadrature the positive and negative errors due to the centring and PA errors. The blue and red regions show the median and 1-sigma percentiles of \Omph{} and \Omkn{} weighted by the average centring plus PA errors, respectively. The 1-sigma percentiles are the associated errors to the length of the slits, and we will refer to them as \dLp{} and \dLm{} with the average error as \dL{}. This is the dominant error source in 3 galaxies of our sample and accounts on average for $\sim 9 \%$ of the relative error.

Notice that the slit with a meassurement closest to the median \Om{} or with the smallest error is usually not the largest one. In Figure \ref{Fig:Omega_Slits} we highlight in purple the slit with the smallest error among the 4 measurements closest to the median. The results presented in tables \ref{Tab:Omega} and \ref{Tab:CR} were obtained using this particular slit length.

\cite{Guo2018} performed tests in a simulated galaxy from \cite{Athanassoula2013}, to study the influence of the slit length to the TW-method. In their tests they show that: (i) the pattern speed converges to the true value of \Om{} at lengths a bit larger than the bar length, (ii) the pattern speed is a monotonically increasing function with the slit length and (iii) large inclination angles ($\gtrsim 60 \degree$) and PA differences between the bar and the disc ($\gtrsim 75 \degree$, and $\lesssim 10 \degree$) can affect the measurement and increase the errors.

In our sample, we observe a convergence of \Om{} in 13 galaxies including our 3 example galaxies (see Figure \ref{Fig:Omega_Slits}). Among the 5 remaining galaxies where \Om{} does not converge, 2 of them have a PA difference between the disc and the bar $<10 \degree$ (manga-7990-12704 and manga-8243-12704), and 1 has a large inclination angle (manga-8317-12704). The non-convergence in the remaining two galaxies (manga-8318-12703 and NGC 5406) is more difficult to explain but could be related to the spiral structure that is covered in the IFU FoV.

While most of our galaxies converge to a value of \Om{}, not all of them do it with a monotonically increasing function. Of the 13 galaxies which do converge, 2 of them show a monotonically decreasing function (manga-8453-12701 and NGC 5205), and 3 show an oscillating function that ends up flattening (manga-8312-12704, manga-8313-9101 and NGC 6497). The variety of shapes could originate from small perturbations on the flux and velocity maps that cannot be captured in simulated galaxies. Another possibility could be the effect of structures rotating with different pattern speeds and the effect they induce in the TW-method.

\subsection{Spatial resolution error}
\label{Sec:PSF_error}

The full-width at half-maximum (FWHM) of the point spread function (PSF) in MaNGA and CALIFA is about $2.5^{\prime \prime}$, or about $\sigma_{PSF} \sim 1.06^{\prime \prime}$ assuming a gaussian profile  \citep{Sanchez2012, Yan2016}.
However, due to the sample selection, MaNGA galaxies are located at higher redshifts. Thus, bars in MaNGA tend to have a shorter projected length, and their light is more spread due to the effect of the PSF. Therefore, the physical resolution is considerably worse in many cases than that of the CALIFA dataset. The mean bar radius before de-projection in our MaNGA sample is about $7.5^{\prime \prime}$, while in our selected CALIFA galaxies is $13.5^{\prime \prime}$. In this section, we examine if a lower spatial resolution may introduce a systematic error.

To test this idea, we increased the effective PSF (hereafter $\sigma_{eff}$) of our CALIFA galaxies, by convolving the stellar flux and velocity maps with Gaussian filters with a standard deviation $\sigma_{g}$. The resulting effective PSF after the convolution is  $\sigma_{eff} = (\sigma_g^2 + \sigma_{PSF}^2)^{0.5}$.  We choose the Gaussian filters such that the resulting $\sigma_{eff}$ had values of 2, 3, 4, and 5. Then we measure \Om{} using the smoothed maps and compare with our results without smoothing. To reduce the effects from convolving around the corners we first extrapolated the border of our maps using the function \code{interpolate replace nans} from the \code{astropy} convolution package, with a gaussian kernel with a standard deviation of 1 \citep{Astropy2018}.

\begin{figure} 
  \centering
    \begin{subfigure}{\linewidth}
    \includegraphics[width=\linewidth]{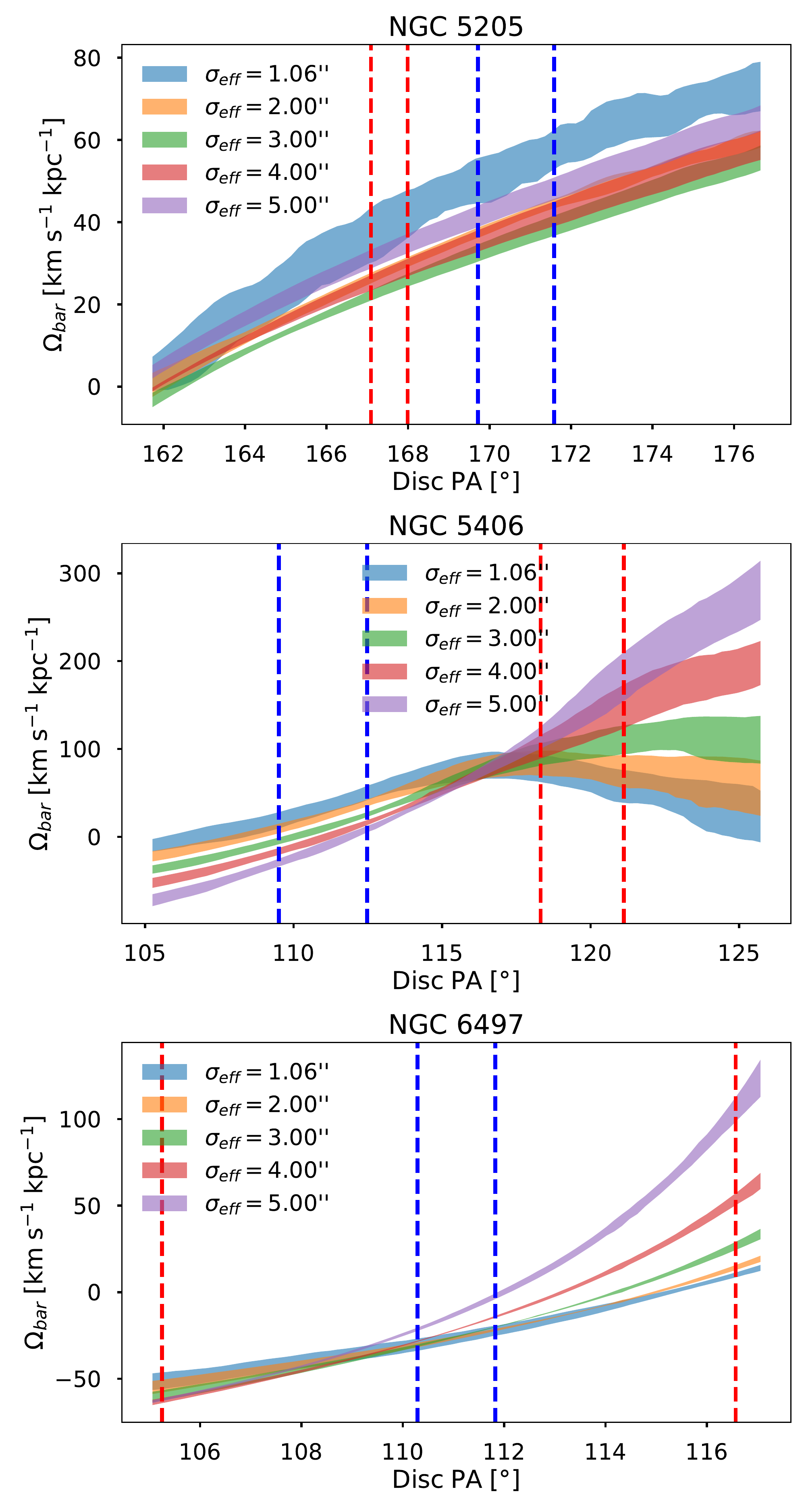}
   	 \end{subfigure}
  \caption{ Same as Figure \ref{Fig:PA-error}, but for our CALIFA sample. The blue and red segmented lines show the location of \PAph{} and \PAkn{} within their respective uncertainties. Each shaded region shows the measurements of \Om{} with a different resolution. In particular, the light-blue region shows the results without smoothing.}
    \label{Fig:Omega_PA_Gauss}
\end{figure}

Similar to Figure \ref{Fig:PA-error}, in Figure \ref{Fig:Omega_PA_Gauss} we show the \Om{} vs PA plots of our CALIFA galaxies, after convolving the maps with different Gaussian filters. The original resolution of the bar does not seem to affect the sensitivity to the change of PSF. NGC 5406 which has the largest projected bar, is the most affected by the convolutions. The least affected galaxy was NGC 5205 where \Omph{} and \Omkn{} changes by $\approx 20 \%$ which remains within a 2-$\sigma$ error of the original measurement. More interestingly, notice the change of the slope between \Om{} and the PA. In the cases of NGC 5406 and NGC 6497, the slope becomes noticeably steeper as we increase $\sigma_{eff}$, making our estimation of \Om{} more sensitive to the PA error. 

In the original paper, \cite{Tremaine1984} argue that the spatial resolution should not bias the pattern speed. This was confirmed by \cite{Zou2019} using mock observations. However, our test suggests that, while the value of \Om{} remains close to the 2-sigma error, the sensitivity to the PA error does increase. This makes sense if we consider that lowering the spatial resolution spreads the underlying substructures to the rest of the galaxy. Thus, we choose to not treat the resolution error as another independent error source, and consider that its effects are implicit in the error estimations we made. Quantifying how much the underling uncertainties propagate requires a more careful analysis that is beyond this paper scope.

\subsection{Bar pattern speed estimation, and the importance of the error treatment}
\label{Sec:Bar_pattern}

All the uncertainty sources we studied arise from the geometric nature of the TW-method. They are different in each galaxy, are asymmetrical, depending on the slits orientation, the light concentration and other galactic structures. We do not observe a significant correlation between the different error sources. For simplicity, we treat them as independent error sources.

Our final estimation of \Om{} is the  weighted by errors obtained from the slit length test. The associated errors ($\delta \Omega_{bar}^+, \delta\Omega_{bar}^-$) are the summing in quadrature of the positive and negative errors separately. In Table \ref{Tab:Omega} we show the resulting \Om{} and the uncertainties due to each error source. The average relative error is shown in parentheses.

The greatest error source is different for each galaxy. Recapitulating what we mentioned in previous sections, the centring error was the dominant error source in 2 galaxies, the PA error in 13 galaxies, and the slits length error in the 3 left. On average they accounted for $\approx 5 \%, 15 \%$ and $9 \%$ of the relative error respectively.

We tested the effects of using a more traditional approach in the error treatment of the \Om{} estimates. We repeated all the tests but assuming we can describe the errors with a Gaussian distribution using the mean and standard deviation, instead of the weighted median and 1-sigma percentiles. Not taking into account the cited asymmetries results in wider errors. The average relative errors related to the centring, PA and slit length increase to $\approx 10 \%$, $19 \%$ and $17 \%$ respectively. This wider distribution of errors in \Om{} consequently affect the estimations of \CR{} and the \R{} parameter, as we explain in the next section.

\begin{table*} 
\centering
\caption{Bar pattern speed and the main error sources for our sample.}
\label{Tab:Omega}
{
\renewcommand{\arraystretch}{1.3}
\begin{tabular}{cccccc}
\hline \hline
 Galaxy &    \Om{} & $\delta \Omega_C \: \left( \frac{\delta \Omega_C}{\Omega_{bar}} \right) $ & $\delta \Omega_{PA} \: \left( \frac{\delta \Omega_{PA}}{\Omega_{bar}} \right)$ & $\delta \Omega_{Len} \: \left( \frac{\delta \Omega_{Len}}{\Omega_{bar}} \right)$  \\
       &   [$\text{km} \: \text{s}^{-1} \: \text{kpc}^{-1}$]  &  [$\text{km} \: \text{s}^{-1} \: \text{kpc}^{-1}$] &   [$\text{km} \: \text{s}^{-1} \: \text{kpc}^{-1}$]  &   [$\text{km} \: \text{s}^{-1} \: \text{kpc}^{-1}$] \\
   (1)  &  (2)  & (3) & (4) & (5) \\        
\hline
  manga-7495-12704 &    $36.7^{+5.7}_{-5.0}$ &  $^{+0.8}_{-0.6} \; \;(0.02)$ &    $^{+5.6}_{-4.0} \; \;(0.13)$ &   $^{+0.9}_{-2.9} \; \;(0.05)$ \\
 manga-7962-12703 &    $22.8^{+1.6}_{-1.9}$ &  $^{+0.3}_{-0.5} \; \;(0.02)$ &    $^{+1.6}_{-1.7} \; \;(0.07)$ &   $^{+0.2}_{-0.7} \; \;(0.02)$ \\
 manga-7990-12704 &    $35.1^{+4.0}_{-7.6}$ &  $^{+0.8}_{-0.3} \; \;(0.02)$ &    $^{+2.6}_{-5.8} \; \;(0.12)$ &   $^{+3.0}_{-4.9} \; \;(0.11)$ \\
  manga-8135-6103 &    $24.9^{+1.8}_{-3.8}$ &  $^{+0.1}_{-0.4} \; \;(0.01)$ &    $^{+1.7}_{-3.7} \; \;(0.11)$ &   $^{+0.8}_{-0.8} \; \;(0.03)$ \\
 manga-8243-12704 &  $36.0^{+18.8}_{-11.7}$ &  $^{+9.2}_{-4.4} \; \;(0.19)$ &   $^{+10.5}_{-5.8} \; \;(0.23)$ &  $^{+12.5}_{-9.1} \; \;(0.30)$ \\
  manga-8256-6101 &  $57.0^{+13.1}_{-18.0}$ &  $^{+2.5}_{-3.2} \; \;(0.05)$ &  $^{+12.8}_{-17.4} \; \;(0.27)$ &   $^{+1.2}_{-3.0} \; \;(0.04)$ \\
  manga-8257-3703 &    $36.7^{+3.5}_{-2.9}$ &  $^{+2.4}_{-1.9} \; \;(0.06)$ &    $^{+2.1}_{-2.1} \; \;(0.06)$ &   $^{+1.5}_{-0.3} \; \;(0.02)$ \\
 manga-8312-12704 &    $13.8^{+8.9}_{-3.9}$ &  $^{+0.4}_{-0.5} \; \;(0.03)$ &    $^{+8.7}_{-3.8} \; \;(0.45)$ &   $^{+1.5}_{-0.6} \; \;(0.08)$ \\
  manga-8313-9101 &    $56.7^{+7.3}_{-6.2}$ &  $^{+1.0}_{-0.6} \; \;(0.01)$ &    $^{+6.8}_{-6.0} \; \;(0.11)$ &   $^{+2.6}_{-1.6} \; \;(0.04)$ \\
 manga-8317-12704 &    $13.9^{+2.3}_{-2.1}$ &  $^{+0.3}_{-0.9} \; \;(0.04)$ &    $^{+1.9}_{-1.4} \; \;(0.12)$ &   $^{+1.2}_{-1.3} \; \;(0.09)$ \\
 manga-8318-12703 &    $29.3^{+5.4}_{-8.4}$ &  $^{+1.9}_{-1.5} \; \;(0.06)$ &    $^{+3.4}_{-5.6} \; \;(0.15)$ &   $^{+3.8}_{-6.1} \; \;(0.17)$ \\
 manga-8341-12704 &    $27.1^{+6.7}_{-7.7}$ &  $^{+3.0}_{-3.0} \; \;(0.11)$ &    $^{+6.0}_{-7.0} \; \;(0.24)$ &   $^{+0.4}_{-1.1} \; \;(0.03)$ \\
  manga-8439-6102 &    $34.2^{+4.2}_{-4.7}$ &  $^{+2.3}_{-2.2} \; \;(0.07)$ &    $^{+3.4}_{-3.9} \; \;(0.11)$ &   $^{+1.0}_{-1.6} \; \;(0.04)$ \\
 manga-8439-12702 &    $29.0^{+5.0}_{-4.7}$ &  $^{+1.5}_{-1.1} \; \;(0.04)$ &    $^{+4.8}_{-3.7} \; \;(0.15)$ &   $^{+0.3}_{-2.6} \; \;(0.05)$ \\
 manga-8453-12701 &   $28.3^{+15.1}_{-3.0}$ &  $^{+0.8}_{-1.8} \; \;(0.05)$ &    $^{+2.4}_{-2.1} \; \;(0.08)$ &  $^{+14.9}_{-1.2} \; \;(0.28)$ \\
          NGC 5205 &    $53.4^{+4.1}_{-5.3}$ &  $^{+1.2}_{-5.0} \; \;(0.06)$ &    $^{+1.5}_{-0.9} \; \;(0.02)$ &   $^{+3.6}_{-1.4} \; \;(0.05)$ \\
          NGC 5406 &   $37.2^{+5.4}_{-12.9}$ &  $^{+1.9}_{-3.6} \; \;(0.07)$ &    $^{+4.5}_{-9.9} \; \;(0.19)$ &   $^{+2.4}_{-7.5} \; \;(0.13)$ \\
          NGC 6497 &    $30.5^{+5.4}_{-5.0}$ &  $^{+2.1}_{-2.2} \; \;(0.07)$ &    $^{+3.7}_{-4.2} \; \;(0.13)$ &   $^{+3.4}_{-1.5} \; \;(0.08)$ \\
\hline
\end{tabular}
}
\caption*{Col.(1) Galaxy ID. Col.(2) Bar pattern speed. Col.(3) Centring error. Col.(4) Position angle error. Col.(5)  Slit length error. In parenthesis we show the average relative error. }
\end{table*}

\subsection{Corotation radius and the \R{} parameter}
\label{Sec:Corotation}

\CR{} is usually estimated using the ratio $V_{flat} / \Omega_{bar}$. This procedure assumes that \CR{} lies in a region where the rotation curve has reached the asymptotic flat regime, which is not necessarily true. This is especially important in late-type galaxies, where the rotation curve tends to be slow-rising \citep[e.g.][]{Kalinova2017}, and \CR{} would be overestimated. 

\begin{figure*}
\centering
	\begin{subfigure}{0.33 \textwidth}
		\includegraphics[width = \textwidth]{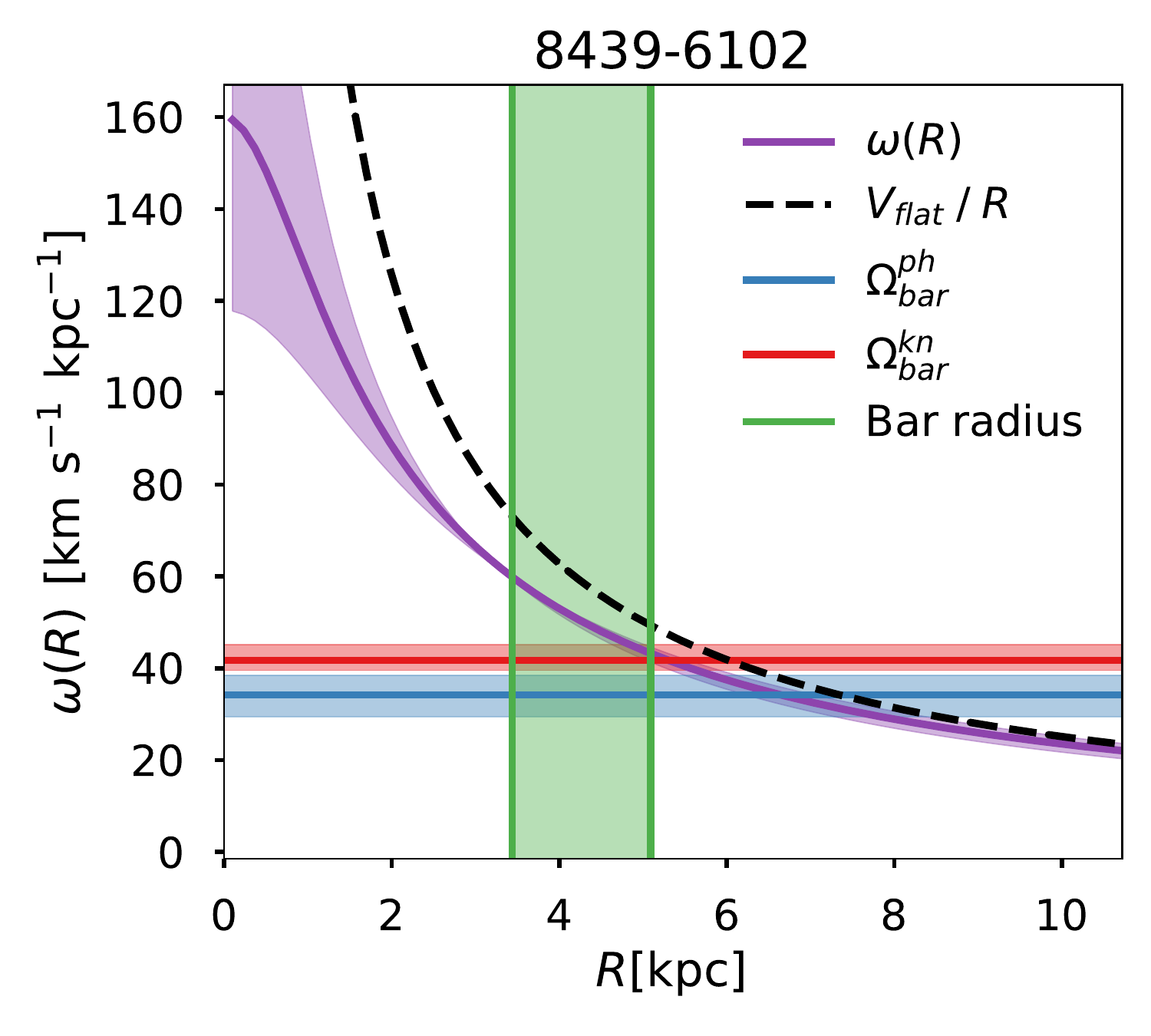}
	\end{subfigure}%
   	\begin{subfigure}{0.33 \textwidth}
		\includegraphics[width = \textwidth]{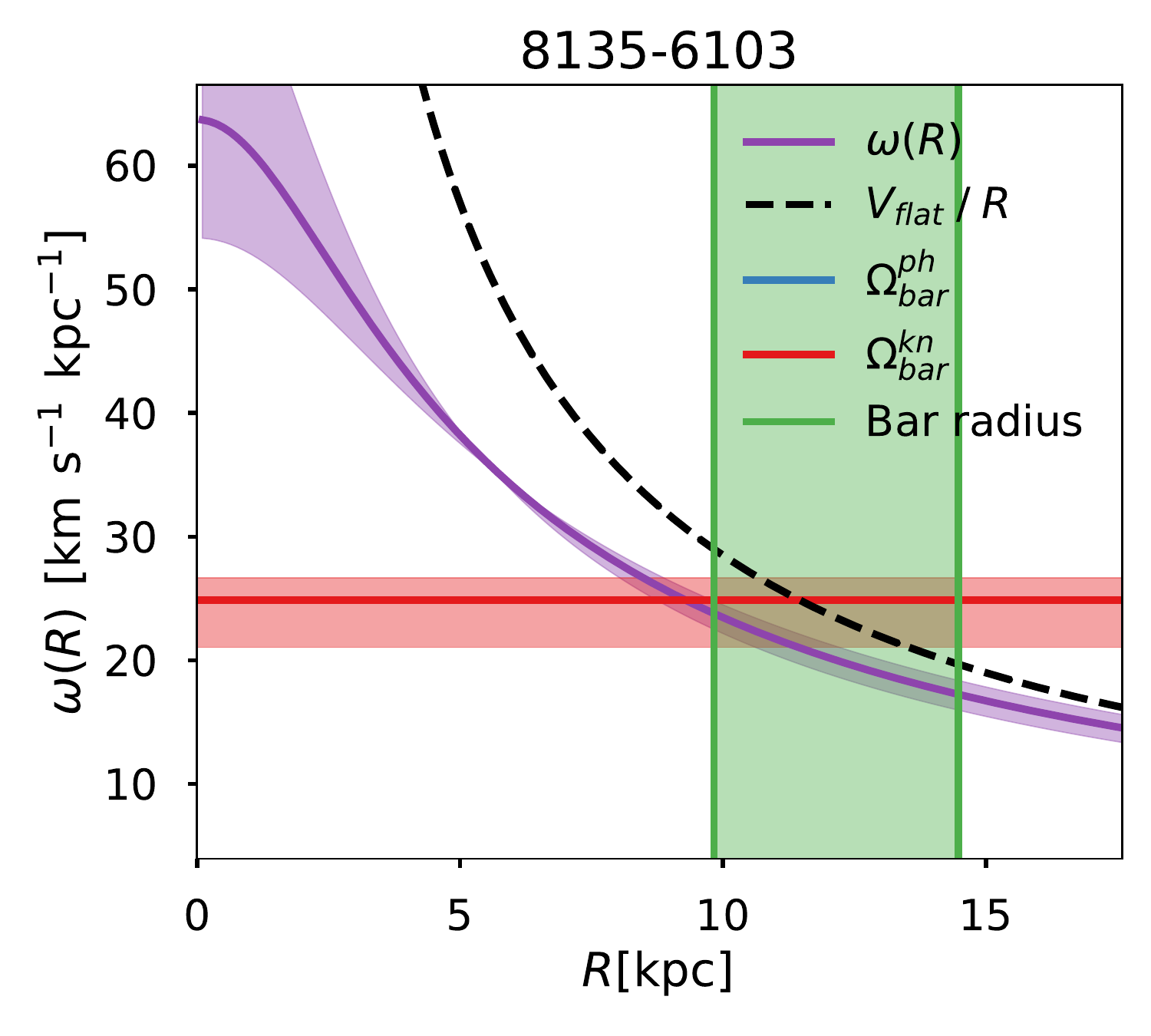}
	\end{subfigure}%
   	\begin{subfigure}{0.33 \textwidth}
		\includegraphics[width = \textwidth]{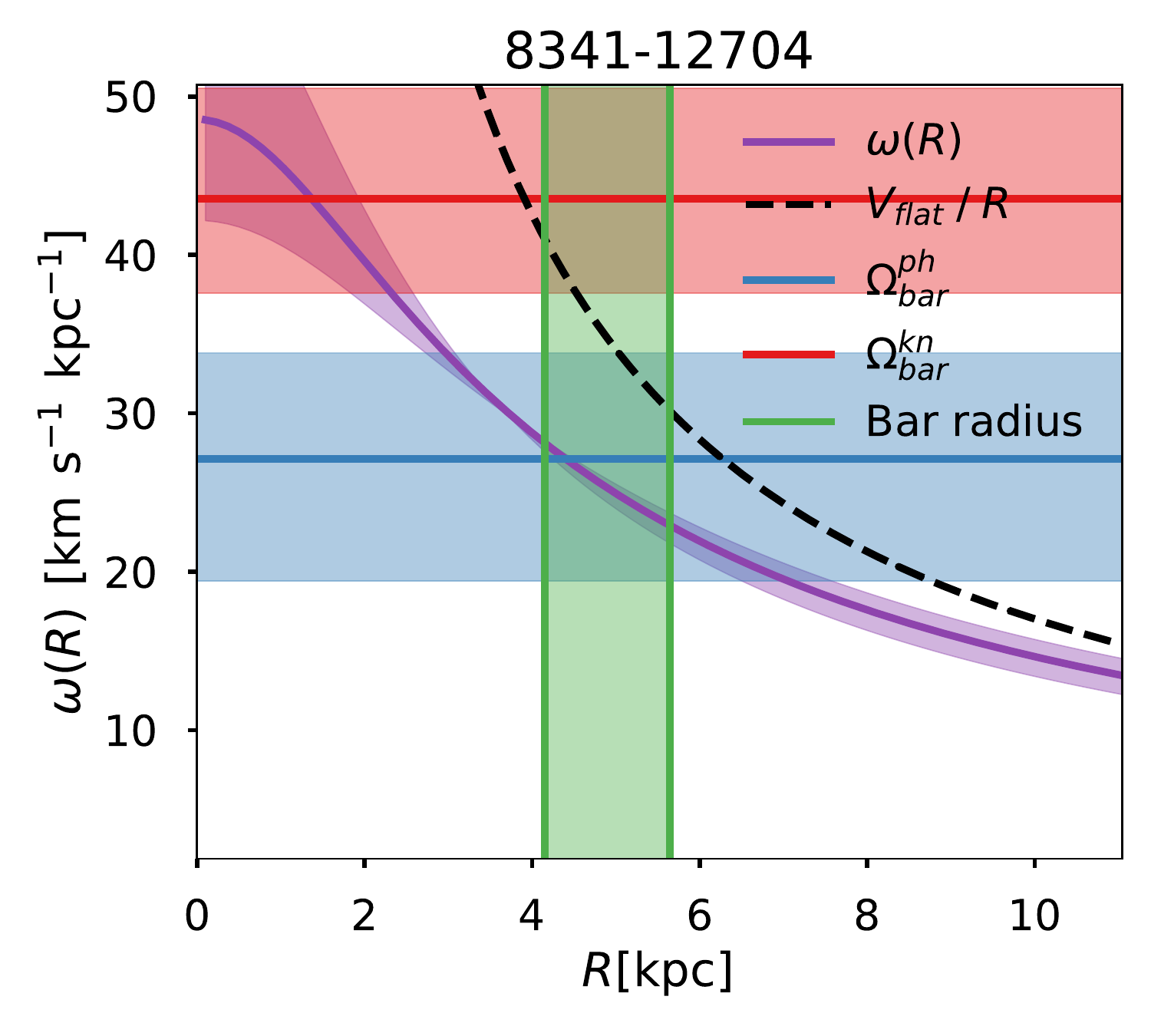}
	\end{subfigure}
    \begin{subfigure}{0.33 \textwidth}
		\includegraphics[width =  \textwidth]{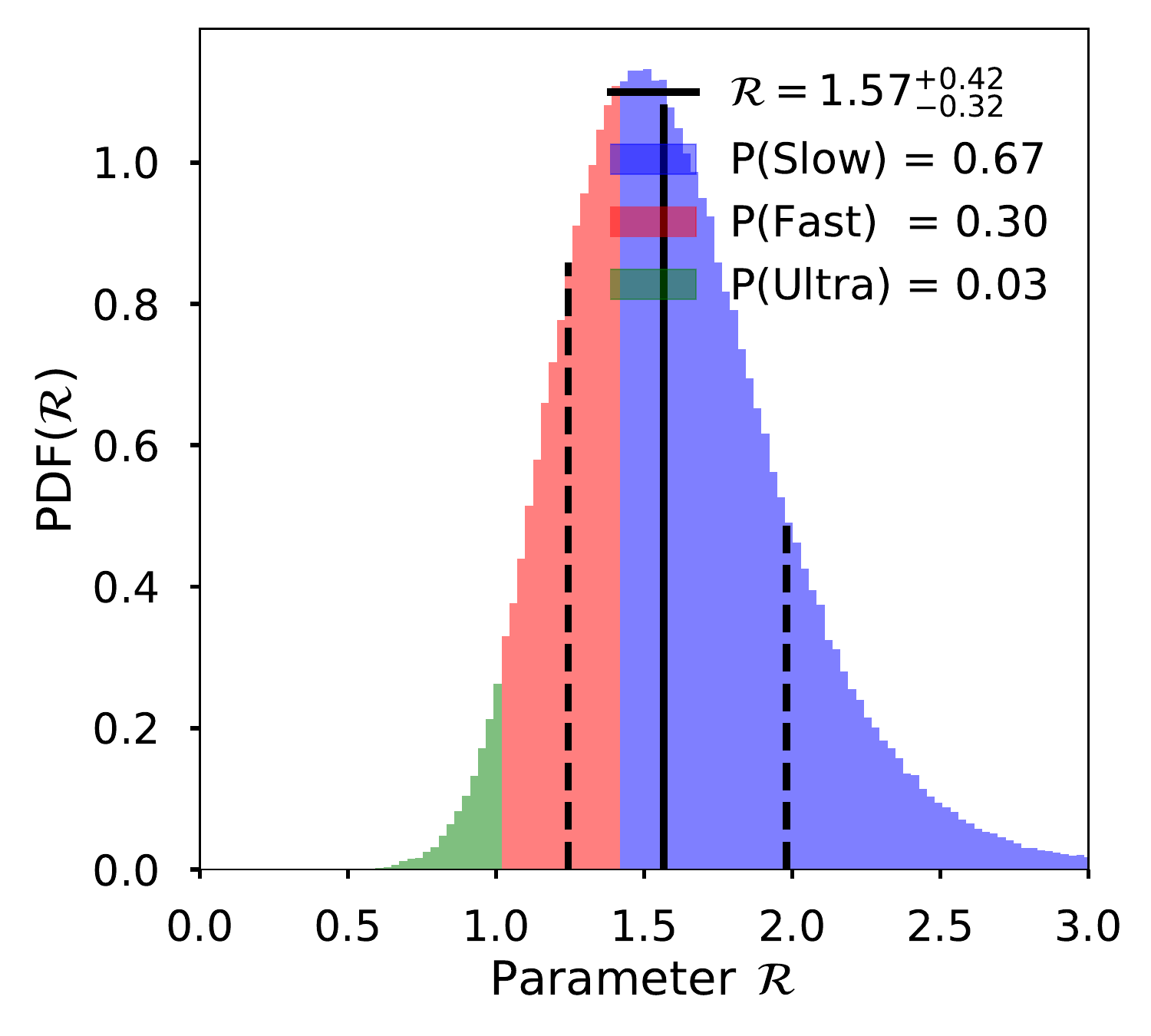} 
    \end{subfigure}%
    \begin{subfigure}{0.33 \textwidth}
		\includegraphics[width =  \textwidth]{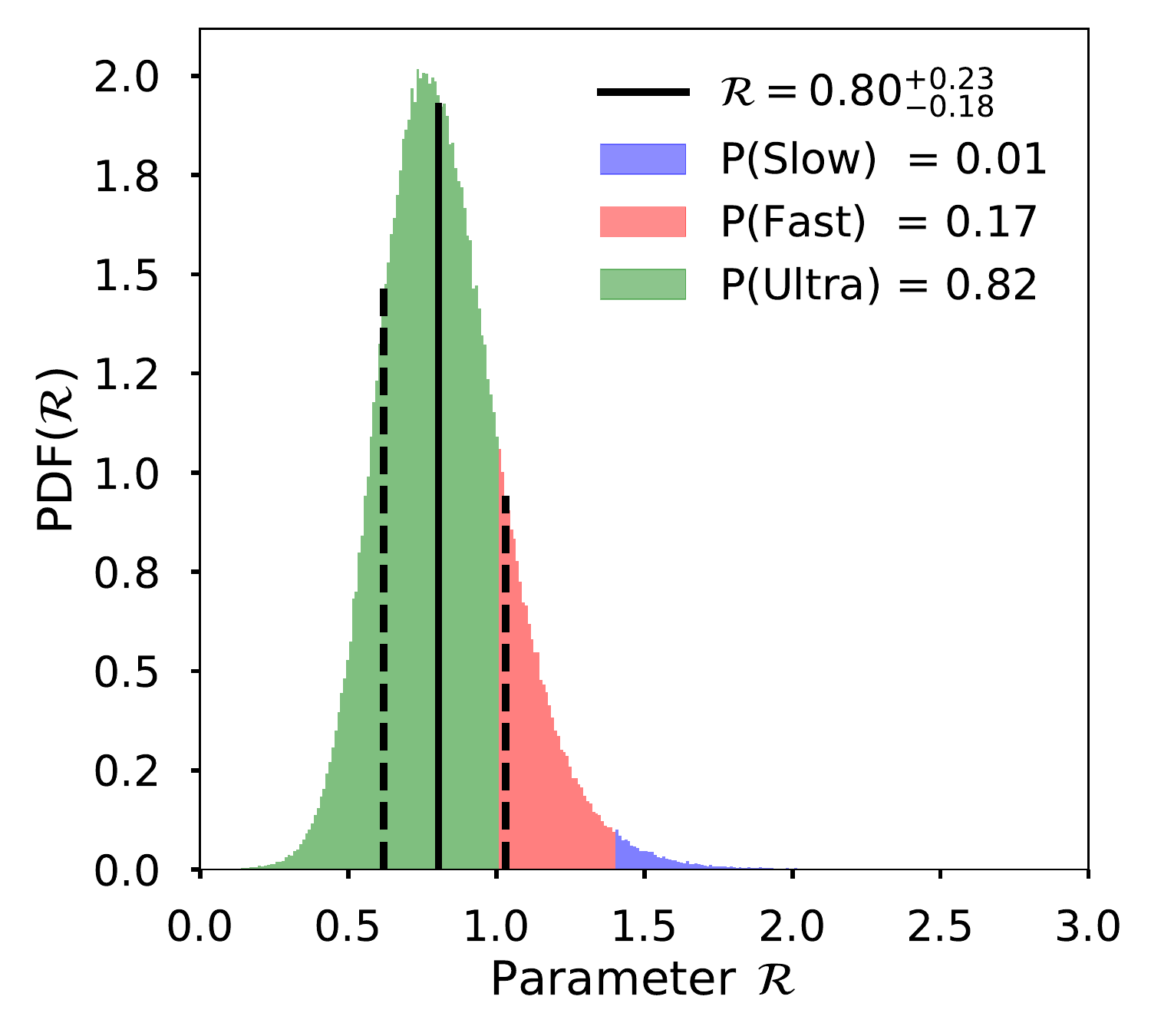} 
    \end{subfigure}%
    \begin{subfigure}{0.33 \textwidth}
		\includegraphics[width =  \textwidth]{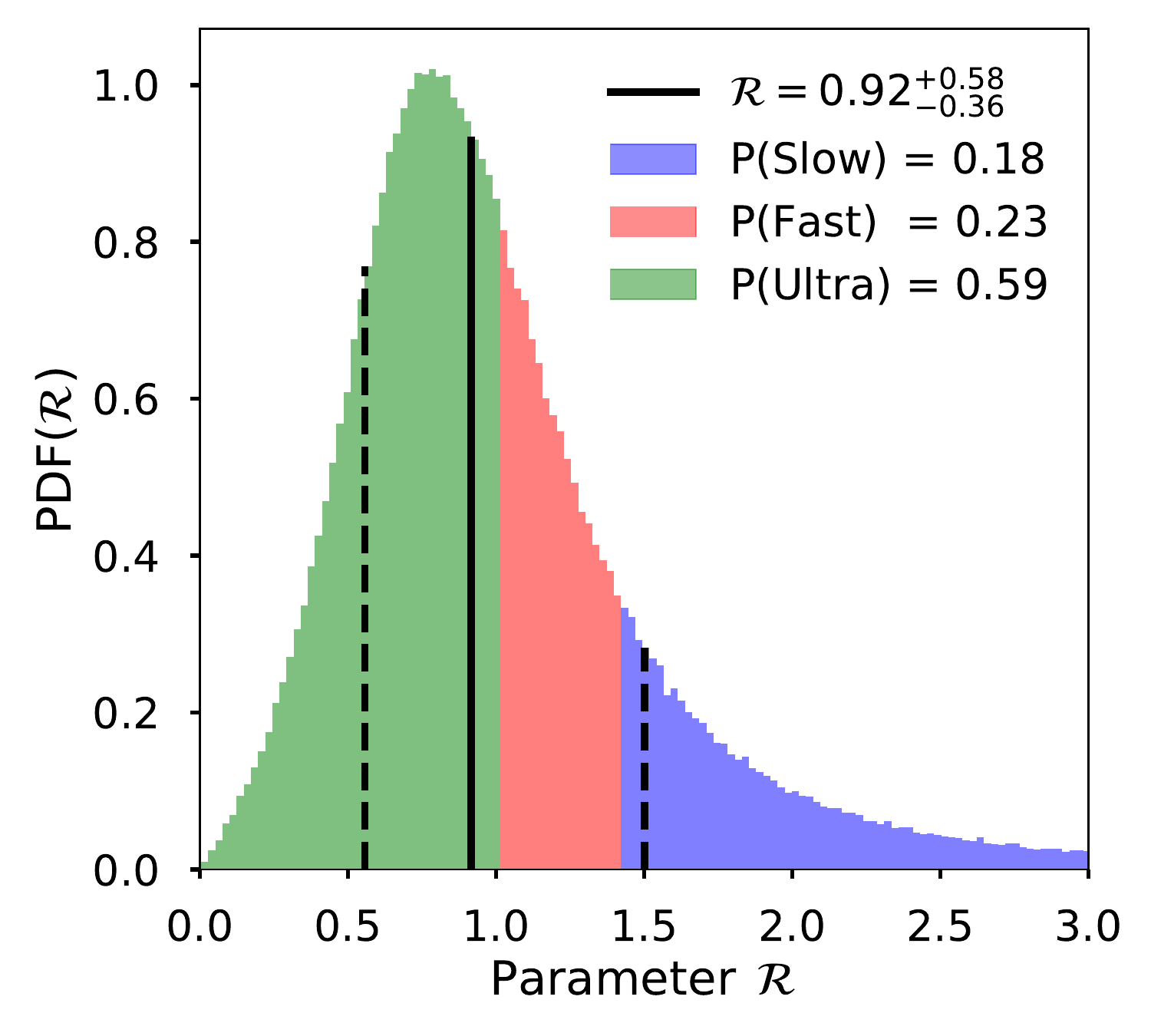}    
    \end{subfigure}   
	\caption{ \textit{Top panels}: Angular velocity curve of our example galaxies obtained from the best \code{Velfit} model. The dashed line shows the curve $V_{flat}/R$. The blue and red regions show the measurements of \Omph{} and \Omkn{}. The green region shows the de-projected bar radius. \textit{Bottom panels}: Probability Distribution Function of the \R{} Parameter. The area under the curve is colored depending on the bar kinematic classification: green for ultra-fast, red for fast and blue for slow. The black solid line shows the median of the distribution, and the dashed lines show the 1-sigma percentiles.}
   	\label{Fig:Ang_curves}
\end{figure*}

In this work, we estimate \CR{} as the intersection between \Om{} and the angular rotation curve (\angcur{}) modelled by \code{Velfit}. In the top panels of Figure \ref{Fig:Ang_curves} we plot both \angcur{} and $V_{flat} / R$ (purple region and dashed line respectively) and compare them with our measurements of \Omph{} and \Omkn{} (blue and red regions) of our example galaxies. We also show the de-projected bar radius (green region). Both manga-8135-6103 and manga-8341-12704 are examples galaxies with slow-rising rotation curves, where the difference in \CR{} is more significant. On average, the relative difference between both methods is $\approx 21 \%$ in our sample.

To estimate \CR{} we performed a MC simulation over the uncertainties of \angcur{} and \Om{} and look for their intersection in each iteration. We modeled the two parameters of the rotation curve using a Gaussian distribution with dispersion equal to their associated errors. For \Om{} we used a log-normal distribution with mean and standard deviation ($\mu$, $\sigma$) equal to

\begin{equation}
    \mu = \sqrt{ \left( \Omega_{bar} + \delta \Omega_{bar}^{+}\right)  \left( \Omega_{bar} - \delta \Omega_{bar}^{-}\right)}
\end{equation}
\begin{equation}
    \sigma = \left(\Omega_{bar} + \delta \Omega_{bar}^{+}\right) / \mu
\end{equation}

These values of $\mu$ and  $\sigma$ produce a log-normal distribution that has the same 1-sigma dispersion as \Om{} with a slightly different mean. This is a good approximation when \Om{} is skewed to the right (\dOmp{} > \dOmm{}). When \Om{} is skewed to the left (\dOmp{} < \dOmm{}), we reflect the distribution (multiplying by $-1$) and translate it back to the same 1-sigma range (by adding  $2\times\Omega_{bar} + \delta \Omega_{bar}^+ - \delta \Omega_{bar}^-$). The reflected distribution can produce some negative values for \Om{}, however in all cases they occur with a frequency $\lesssim 1 \%$, so their effect should be negligible.

Similarly, to estimate the value of the \R{} parameter, we performed a second MC simulation by dividing the resulting elements of \CR{} with a random bar radius modelled with a uniform distribution over the range of $\left( \mathrm{R}_{bar}^{dep,1}, \mathrm{R}_{bar}^{dep,2} \right) $. The bottom panels of Figure \ref{Fig:Ang_curves} show the resulting Probability Distribution Function (PDF) of the \R{} parameter. The area under the curve is coloured depending on the bar classification. The black solid line shows the median value, and the dashed lines show the 1-sigma percentiles of the distribution. In Table \ref{Tab:CR} we show our measurements for \CR{}, \R{}, and the probabilities for each bar classification. 

As we mention in Section \ref{Sec:Rotation_curve} for the galaxy manga-7990-12704 we used the stellar velocity map to derive $V(R)$ without performing an asymmetric drift correction. Thus, we may be underestimating the real \CR{} and the parameter \R{} in this particular ultrafast galaxy.

\begin{table*}
\caption{Corrotation radius, parameter \R{}, and bar classification probabilities of our sample}
\label{Tab:CR}
\centering
{
\renewcommand{\arraystretch}{1.3}
\begin{tabular}{cccccc}
\hline \hline
 Galaxy & \CR{} &  \R & P(Slow) & P(Fast) & P(Ultra) \\
        &  [kpc] &  &  &  &  \\
  (1) & (2) & (3) & (4) & (5) & (6)  \\
\hline
 manga-7495-12704 &   $5.52^{+1.10}_{-0.97}$ &  $1.16^{+0.30}_{-0.24}$ &     0.21 &     0.52 &      0.27 \\
 manga-7962-12703 &  $13.23^{+2.14}_{-2.05}$ &  $0.89^{+0.15}_{-0.14}$ &     0.00 &     0.23 &      0.77 \\
 manga-7990-12704 &   $5.00^{+2.67}_{-2.20}$ &  $0.76^{+0.41}_{-0.33}$ &     0.07 &     0.18 &      0.74 \\
  manga-8135-6103 &   $9.74^{+2.35}_{-2.05}$ &  $0.80^{+0.23}_{-0.18}$ &     0.01 &     0.17 &      0.82 \\
 manga-8243-12704 &   $4.31^{+3.23}_{-2.20}$ &  $1.39^{+1.07}_{-0.72}$ &     0.49 &     0.20 &      0.31 \\
  manga-8256-6101 &   $4.46^{+2.46}_{-1.46}$ &  $0.98^{+0.55}_{-0.33}$ &     0.20 &     0.27 &      0.53 \\
  manga-8257-3703 &   $5.69^{+0.79}_{-0.74}$ &  $1.42^{+0.30}_{-0.24}$ &     0.52 &     0.45 &      0.03 \\
 manga-8312-12704 &  $11.06^{+6.13}_{-4.22}$ &  $1.75^{+0.99}_{-0.68}$ &     0.68 &     0.19 &      0.13 \\
  manga-8313-9101 &   $4.24^{+0.70}_{-0.64}$ &  $1.26^{+0.23}_{-0.21}$ &     0.26 &     0.63 &      0.11 \\
 manga-8317-12704 &  $20.87^{+3.64}_{-3.19}$ &  $1.46^{+0.27}_{-0.24}$ &     0.58 &     0.40 &      0.02 \\
 manga-8318-12703 &   $9.18^{+3.46}_{-1.72}$ &  $1.44^{+0.56}_{-0.30}$ &     0.53 &     0.42 &      0.05 \\
 manga-8341-12704 &   $4.47^{+2.79}_{-1.72}$ &  $0.92^{+0.58}_{-0.36}$ &     0.18 &     0.23 &      0.59 \\
  manga-8439-6102 &   $6.64^{+1.49}_{-1.19}$ &  $1.57^{+0.42}_{-0.32}$ &     0.67 &     0.30 &      0.03 \\
 manga-8439-12702 &   $8.82^{+1.88}_{-1.58}$ &  $1.49^{+0.33}_{-0.27}$ &     0.62 &     0.35 &      0.03 \\
 manga-8453-12701 &   $3.83^{+1.93}_{-1.58}$ &  $1.08^{+0.55}_{-0.45}$ &     0.27 &     0.29 &      0.44 \\
          NGC5205 &   $2.79^{+0.56}_{-0.53}$ &  $1.03^{+0.21}_{-0.20}$ &     0.05 &     0.50 &      0.46 \\
          NGC5406 &   $7.41^{+3.25}_{-1.61}$ &  $0.74^{+0.33}_{-0.17}$ &     0.06 &     0.13 &      0.81 \\
          NGC6497 &   $8.68^{+2.17}_{-1.90}$ &  $1.08^{+0.31}_{-0.25}$ &     0.14 &     0.46 &      0.40 \\
\hline
\end{tabular}
}
\caption*{Col.(1) Galaxy ID. Col.(2) Corotation radius. Col.(3) Parameter  \R. Col.(4) Slow bar probability. Col.(5) Fast bar probability. Col.(6) Ultrafast bar probability }
\end{table*}

\begin{figure}
\centering
	\includegraphics[width = \linewidth]{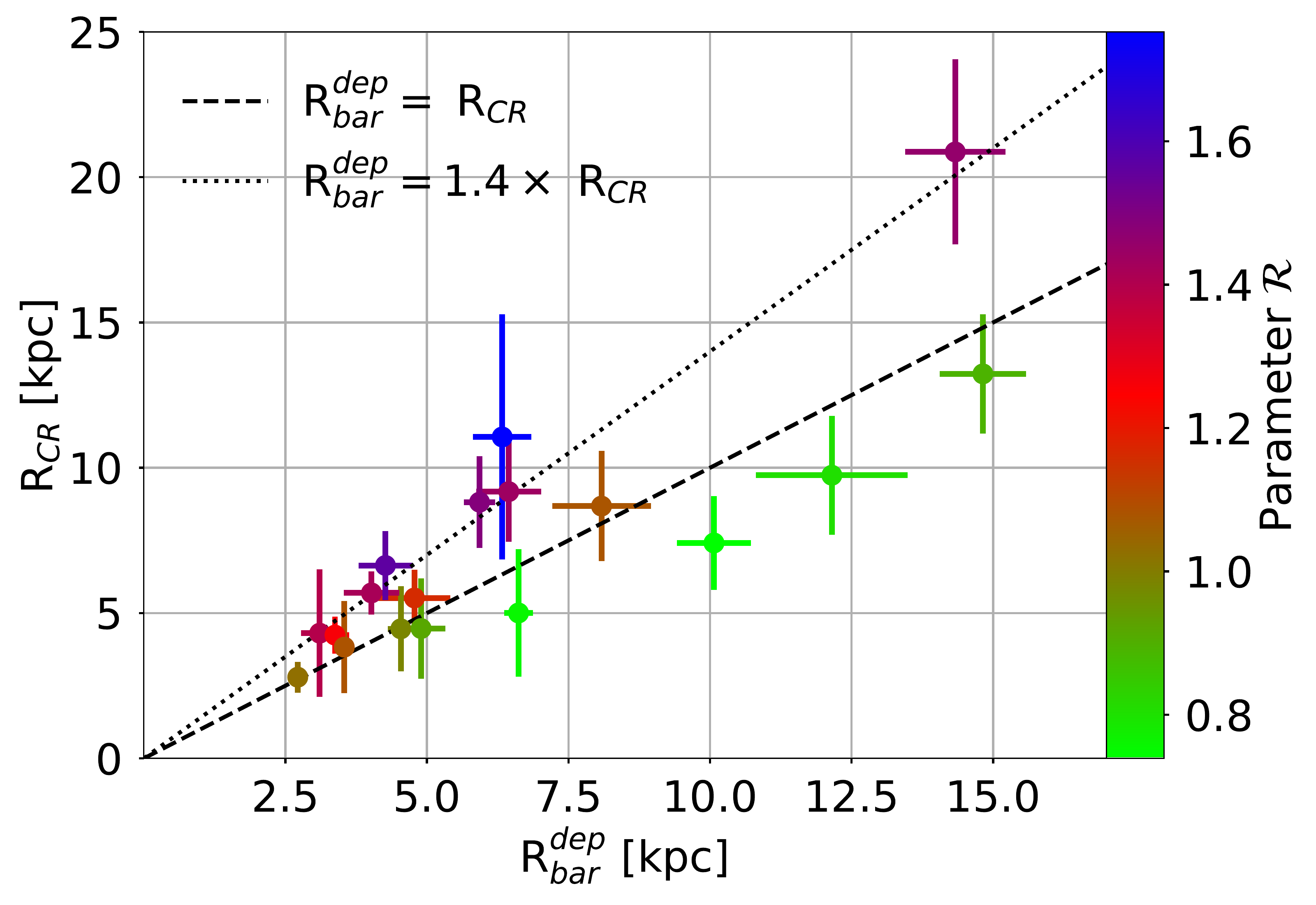}
	\caption{ \CR{} va \Rbar{} of our sample. The colour code shows the median value of the probability distribution function of the  parameter \R{}. The Spearmann correlation coefficient of the relation is \Spear{0.85}}
   	\label{Fig:CR_Rbar}
\end{figure}

According to the median value of \R{} our sample is composed of 6 slow, 6 fast and 6 ultrafast bars. Using the most probable classification, we have 7 slow, 4 fast and 7 ultrafast bars. 4 galaxies are classified with probabilities greater than 1-sigma (P $> 0.683$), all of them as ultrafast bars: manga-7962-12703, manga-7990-12703, manga-8135-6103, and NGC 5406. Figure \ref{Fig:CR_Rbar} shows the \CR{} vs \Rbar{} plot for our sample galaxies, colour-coded by their \R{} parameter.

In general, the resulting PDF distribution of \CR{} and \R{} is always skewed to the right (the positive error is always larger than the negative error). This is expected, from the declining convex shape of \angcur{} and the impossibility of having negative values. This right-skewness could explain the observed frequency of ultrafast rotating bars: the larger the errors in \Om{} result in wider distributions in \CR{} and \R{} with smaller median values and larger tails to the right. Changing the treatment of the errors confirms this prediction. Using improperly symmetrical Gaussian distributions for the errors not only results in wider distributions for \Om{} but also in \CR{} and \R{} parameters. The resulting bar classifications based on probabilities using the improper treatment is 5 slow, 5 fast and 8 ultrafast bars. We expect that further improvements in the treatment of errors could reduce the frequency of the observed ultrafast bars.

\begin{figure}
\centering
	\includegraphics[width = \linewidth]{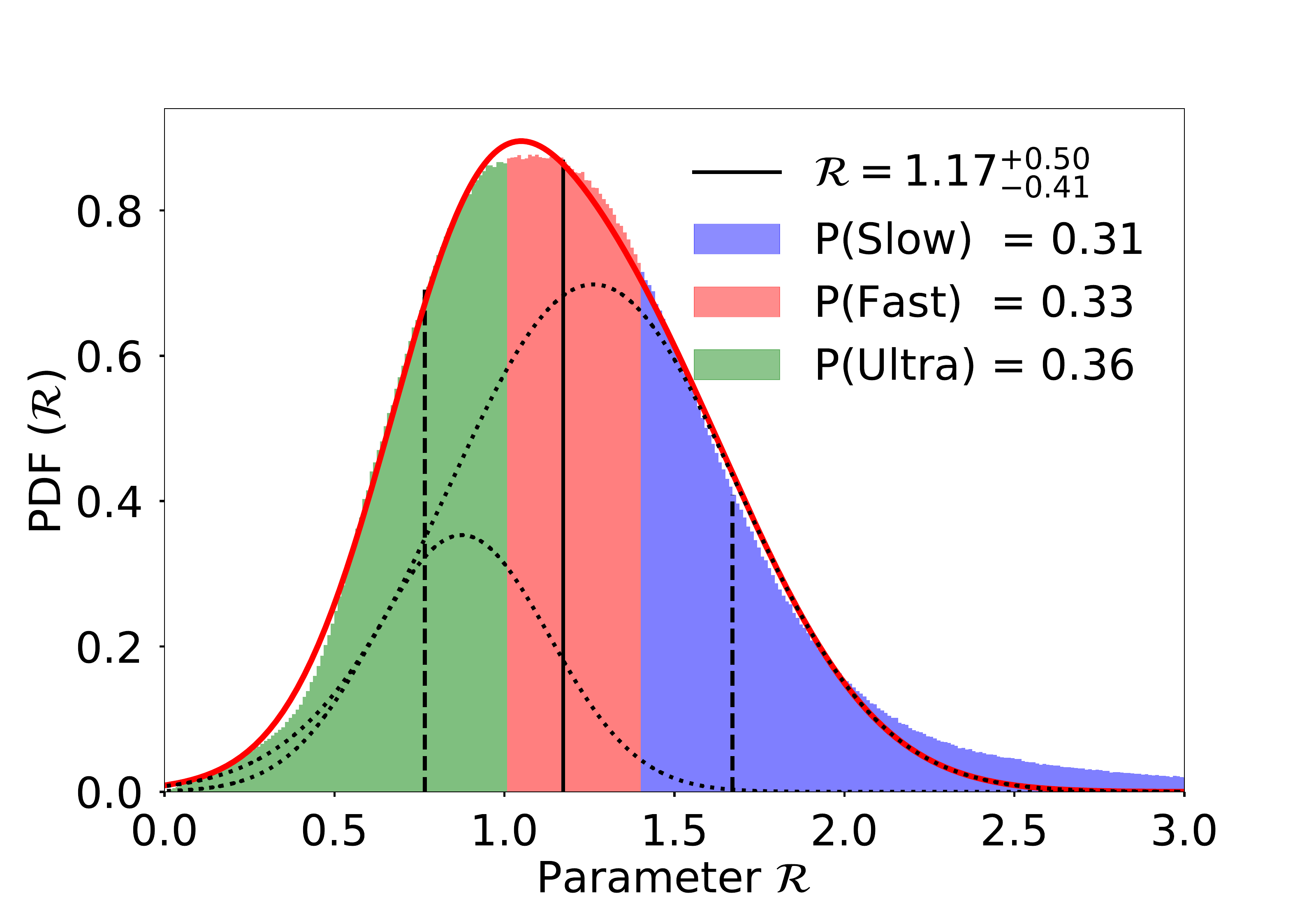}
	\caption{ The parameter \R{} probability distribution function of all our sample added together. Same as Figure \ref{Fig:Ang_curves}, the colour denotes the bar classification. The red line shows the best fit using the sum of two Gaussian distributions (shown with the dotted curves).}
   	\label{Fig:PDF_Total}
\end{figure}

In Figure \ref{Fig:PDF_Total} we show the PDF of \R{} of all our sample added together. It is interesting to notice that the probabilities are distributed almost equally between all bar classifications. We used two Gaussian distribution to model the resulting PDF, which are shown as the dotted curves. The amplitude, centre and standard deviation of these Gaussian distributions are (0.35, 0.87, 0.26) and (0.70, 1.26, 0.42). Other models (single Gaussian or lognormal) do not fit the distribution quite as well. It is not clear if the requirement of two Gaussian curves is a signal of the bars intrinsic evolution or just the result of having a small sample with great individual uncertainties. 

\section{Discussion}
\label{Sec:Discussion}

\subsection{The effect of the uncertainties}

The dominant error source in our sample comes from the PA estimates accounting on average $\approx 15 \%$ of relative error. This is in agreement with the results by  \cite{Debattista2003}. This error is especially important in galaxies were the \Om{} vs PA curve tends to an asymptotic behaviour, which occurs when the slits are oriented close to the major or minor axis of the disc. Naturally, galaxies where the PA is uncertain, are more sensitive to this error. We used prior knowledge of \PAkn{} to further constrain \PAph{} as well as deeper images from the DESI legacy survey. Other important systematic contributions to the error in PA come from the presence of spiral arms, and other non-axisymmetric structures in the outer disc. A promising approach to further constrain the disc PA could be taking into account how the \PAkn{} changes with the radius, due to bars and warped discs \citep{Krajnovic2006, Krajnovic2011, Stark2018} or the usage of Bayesian methods to further constrain \PAph{} given a set of priors obtained from the kinematics.

The slits length contribution to the error appears to be at least as important as the PA error accounting on average $\approx 9 \%$ of relative error. \cite{Guo2018} studied this error using  a simulated galaxy and found that the pattern speed as a function of the slit length is generally a monotonically increasing function that converges to the real value. Their test also shows that the slit length error is sensitive to the difference between the PA of the disc and the bar, and the inclination angle of the disc.
In our tests, we found that most of the galaxies (13/18) do converge to a value of \Om{} however, there are some details that probably are not being captured, in the simulations. For example, in some galaxies, the shape of \Om{} as a function of the slit length is a monotonically decreasing or oscillatory function. This could be an effect due to the spiral structure, or even different patterns in the disc.

We confirmed that the centring error is almost negligible for the TW-method in our sample, accounting for 5 \% of relative error. However, the centring proved to be an important error source in 2 galaxies of our sample, with relative errors $> 10 \%$. This error may be caused by an excess of luminosity in the central regions of these galaxies, affecting the weights in \Xint{} and \Vint{}. 

We tested the effect of changing the effective PSF by smoothing the stellar flux and velocity maps with a Gaussian filter in our CALIFA sub-sample. The relative size of the bar with respect the FoV does not seem to affect the estimation of \Om{}. The most important effect we observe is an increase in the sensitivity to the PA error (see Figure \ref{Fig:Omega_PA_Gauss}). Though a more detailed analysis is needed, it is important to notice the role of good spatial resolution for more accurate measurements of \Om{}. 

The main goal in this work is to identify different error sources in the TW-method and quantify how much they are affecting the final measurement of \Om{}. For simplicity, we assumed they can be treated independently and performed the tests accordingly. We did not observe a significant correlation between the error sources, verifying their independence.
We added the positive and negative errors in quadrature separately. This last assumption could be biasing some results, especially if there are some galaxies where the error sources are indeed correlated. To alleviate this problem, we suggest taking into account the 3 error sources discussed in this paper in the same MC realization. Doing this should also provide the real distribution of \Om{}, which can be used to estimate \CR{} accurately.

The error treatment is also important. Simplifying the analysis using symmetrical errors and Gaussian assumptions result in wider distributions in \Om{}, \CR{} and \R{}. Given the declining convex shape of the angular velocity curve, and the impossibility of having negative values, these quantities should always have a right-skewed distribution. The combination of wider errors and the right-skewness results in an increased frequency of ultrafast rotating bars. Better error treatments, combined with observations with high spatial resolution could resolve the ultrafast bar controversy.

\subsection{Comparing with previous works}

In this work we re-analyze a sub-sample of 3 CALIFA galaxies explored by \cite{Aguerri2015} and 10 MaNGA galaxies we have in common with \cite{Guo2018}. Although we are starting from the same raw data, notice that there are several differences in the procedure for estimating the TW-integrals. For instance, both authors used light and mass weights. The luminous weights were obtained by summing the flux from each spectrum of the datacube in the wavelength range from $4500 \; \text{\AA}$ to $4650 \; \text{\AA}$.  The mass weights were obtained from a stellar population synthesis. For the measurement of \Vint{} \cite{Aguerri2015} used two different approaches: (1) summing directly the velocity inside the pseudo-slits and (2) summing the raw spectra inside the pseudo-slits into one single integrated spectrum, which is analyzed using penalized pixel-fitting method (pPXF) \citep{Cappellari2004}. Since there is no significant difference in the values of \Om{} between these two approaches, \cite{Guo2018} used the former approach. In comparison, we compute \Xint{} and \Vint{} by summing directly over the stellar surface brightness and velocity maps obtained from the data products of \code{Pipe3D}. 

Both works report 4 measurements for \Om{}. To simplify the discussion we looked for the results that are more similar to ours. To find the best comparison, we assumed the tails of the distribution of \Om{} are Gaussian. Then, using the Student's t-test, we found that the best agreement occurs between our photometric PA results (\Omph{}) and the light-weighted kinematic PA results of \cite{Guo2018} and the mass-weighted, velocity sum results from \cite{Aguerri2015}.

The difference in the PA between \cite{Guo2018} and us could be due to the different criteria used to measure the \PAph{}. In this work, we used isophotes estimated on the legacy DESI r-band images, that trace better the outer regions of the galaxy compared to the SDSS r-band images. Most important, however, is the choice of the flat region in the PA profile which is somewhat arbitrary in various galaxies, but that in our case is always estimated within a region 
with a S/N of at least 3. We tried to double-check our measurements by visual inspection of the galaxy and a comparison with the \PAkn{}. In conflicting cases, we used the PA that better resembles the \PAkn{}.
On average, the difference between \cite{Guo2018} photometric PA and ours is $4.4 \degree$. In contrast, the difference between our \PAph{} and their \PAkn{} is $2.9 \degree$. On the other hand, a comparison of our results and those from \cite{Aguerri2015} are in agreement if we consider their mass-weighted results. This may be due to the limited overlapping sample (3 galaxies). 

Another important difference is the procedure used to derive \CR{}. Both authors estimate the corotation radius as: $R_{CR}= V_{flat}/ \Omega_{bar}$. This procedure assumes that corotation is in the flat region of the rotation curve, which is not necessarily true for all galaxies. To measure $V_{flat}$ \cite{Aguerri2015} uses the stellar-streaming velocity along the disc major axis and correct for the asymmetric drift. On the other hand, \cite{Guo2018} estimate $V_{flat}$ from the total mass of the Jeans Anisotropic modeling (JAM) \citep{Cappellari2008}.

In this work, we derived \CR{} by performing a MC simulation over the uncertainties of \Om{} and \angcur{} and look for their intersections, as we explained in Section \ref{Sec:Corotation}. In Figure \ref{Fig:Ang_curves} we show the comparison of $\omega(R)$ and $V_{flat} / R$ in our example galaxies. Using $V_{flat} / \Omega_{bar}$ overestimates \CR{} in galaxies where the rotation curve rises slowly which is the case of our example galaxies manga-8135-6103 and manga-8341-12704. In our sample, the average relative difference in \CR{} between both approaches is $\approx 21 \%$.

The third difference is the number of pseudo slits used.  \cite{Aguerri2015} and \cite{Guo2018} used three to five pseudo-slits separated by a minimum interval of 1$''$ spaxels to avoid overlapped pixels. In contrast, we tried to use as many slits as we could inside the bar region without overlapping pixels as described in \ref{Sec:Number_Sep_Leng}. However, using a different number of slits should only affect the fitting error of the straight line in the \Vint{} vs \Xint{} diagram. This does not affect the results since the fitting error is small compared with the other error sources.

\subsection{Relationships between the bar measurements with other galactic parameters}
\label{Sec:Relations}

Various works have tried to found correlations between the bar properties and different galactic parameters including the Hubble morphological type \citep{Rautiainen2008}, the gas content, \citep{Masters2012, Athanassoula2013} and the stellar mass \citep{Erwin2018}. In this section we present preliminary correlations we observe in our data. Since our sample is small in number and the errors are large, these correlations should be considered only as exploratory in nature and as hints of any possible underlying bar evolution. Given the nature of the errors, the significance of the correlations is not the same when compared with the results from \cite{Aguerri2015} and \cite{Guo2018}. This complicates a direct comparison. 

For example, we observe a strong correlation (Spearmann correlation coefficient of \Spear{0.85})  between \Rdep{} and \CR{}, which can be seen in Figure \ref{Fig:CR_Rbar}. This correlation arises from the high probabilities of fast and ultrafast bars we found in our sample. The works from \cite{Guo2018} and \cite{Aguerri2015} both get \Spear{0.64}. 

We observe a strong anti-correlation (\Spear{-0.68}) between \Om{} and \CR{} which is expected from the definition of \CR{} and the declining shape of the angular velocity curves. This strong anti-correlation is also present in \cite{Guo2018} and \cite{Aguerri2015} (\Spear{-0.91} and \Spear{-0.72} respectively).  

\begin{figure} 
  \centering
    \includegraphics[width=\linewidth]{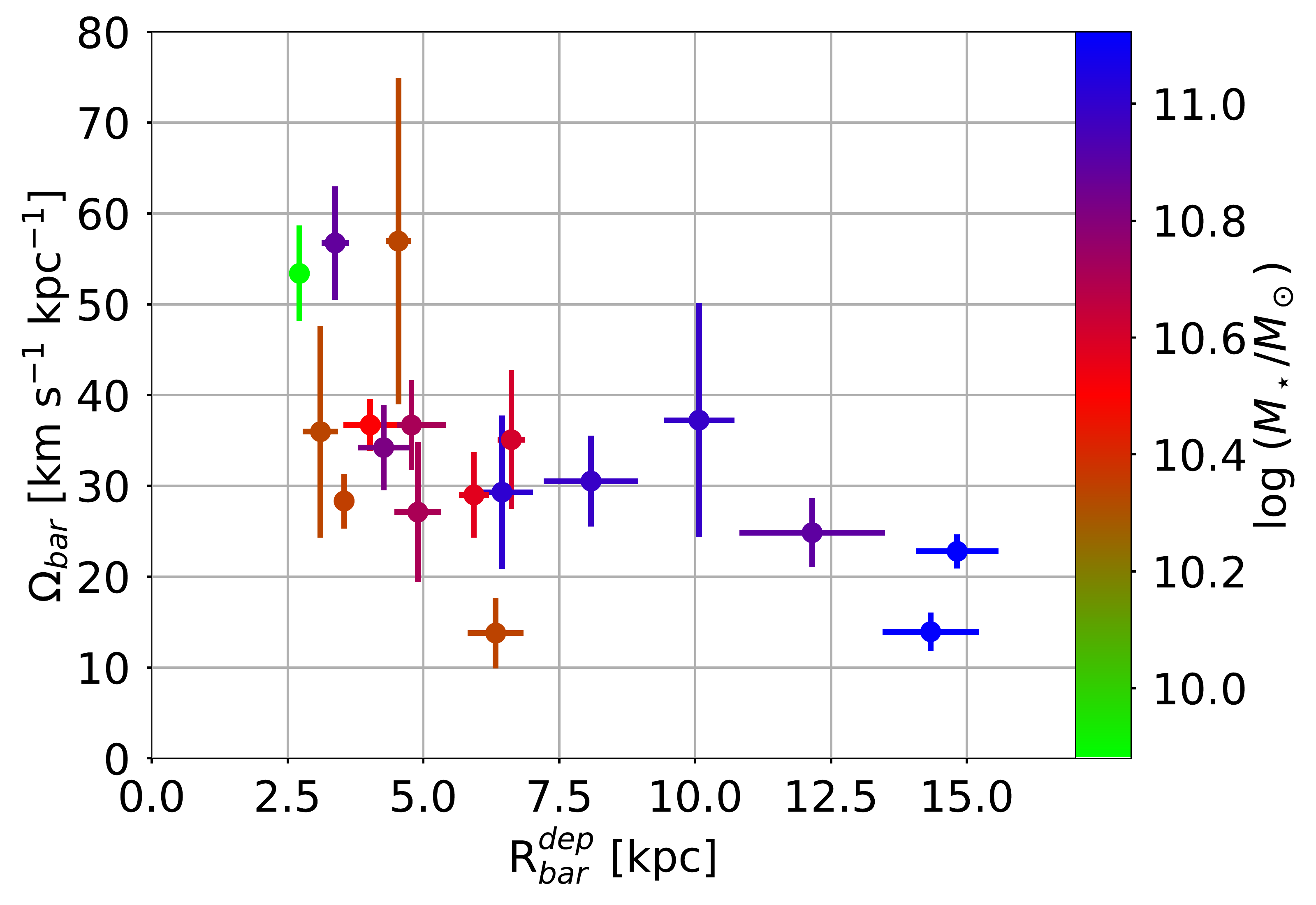}
   	 \caption{Bar pattern speed vs. deprojected bar radius of our sample. The colour code shows the stellar mass of the galaxy. The Spearman correlation coefficient of the relation is \Spear{-0.58} }
  \label{Fig:Omega-Rbar}
\end{figure}

Another anti-correlation (\Spear{-0.58}) is observed between \Om{} and \Rdep{}, suggesting that longer bars tend to rotate at lower pattern speeds. This is consistent with the predictions from N-body simulations, where the bar tends to grow in size and slows down with time \citep[e.g.,][]{Weinberg1985, Athanassoula2013, Martinez-Valpuesta2006}. In Figure \ref{Fig:Omega-Rbar} we show the relationship between these two parameters. Our results are similar to those presented by \citet[][see their Figure 6]{Font2017} where large bars can only rotate with low \Om{}. We also confirm their finding regarding small bars (\Rbar{} $\sim 3 kpc$) with high \Om{} being present in intermediate-mass galaxies. Notice that we have an unusual number of large bars with lengths higher than 8 kpc and high stellar mass ($\log M /M_\odot > 10.8 $). Although that is not intended from our sample selection criteria, this could be biasing some of the cited relationships. The works by \cite{Guo2018} and \cite{Aguerri2015} find weaker anti-correlations (\Spear{-0.55} and \Spear{-0.34} respectively).

We do not find any correlation between the \R{} parameter and the total stellar mass nor the the Hubble type (\Spear{-0.15} and \Spear{-0.10} respectively), confirming the results by \cite{Aguerri2015}. However we do find a weak correlation between \R{} and the molecular gas estimation $M_{gas}$ (\Spear{0.43}). However this relation is almost non-existent in \cite{Guo2018} and \cite{Aguerri2015} (\Spear{0.13} and \Spear{0.20} respectively).

\begin{figure} 
  \centering
    \includegraphics[width=\linewidth]{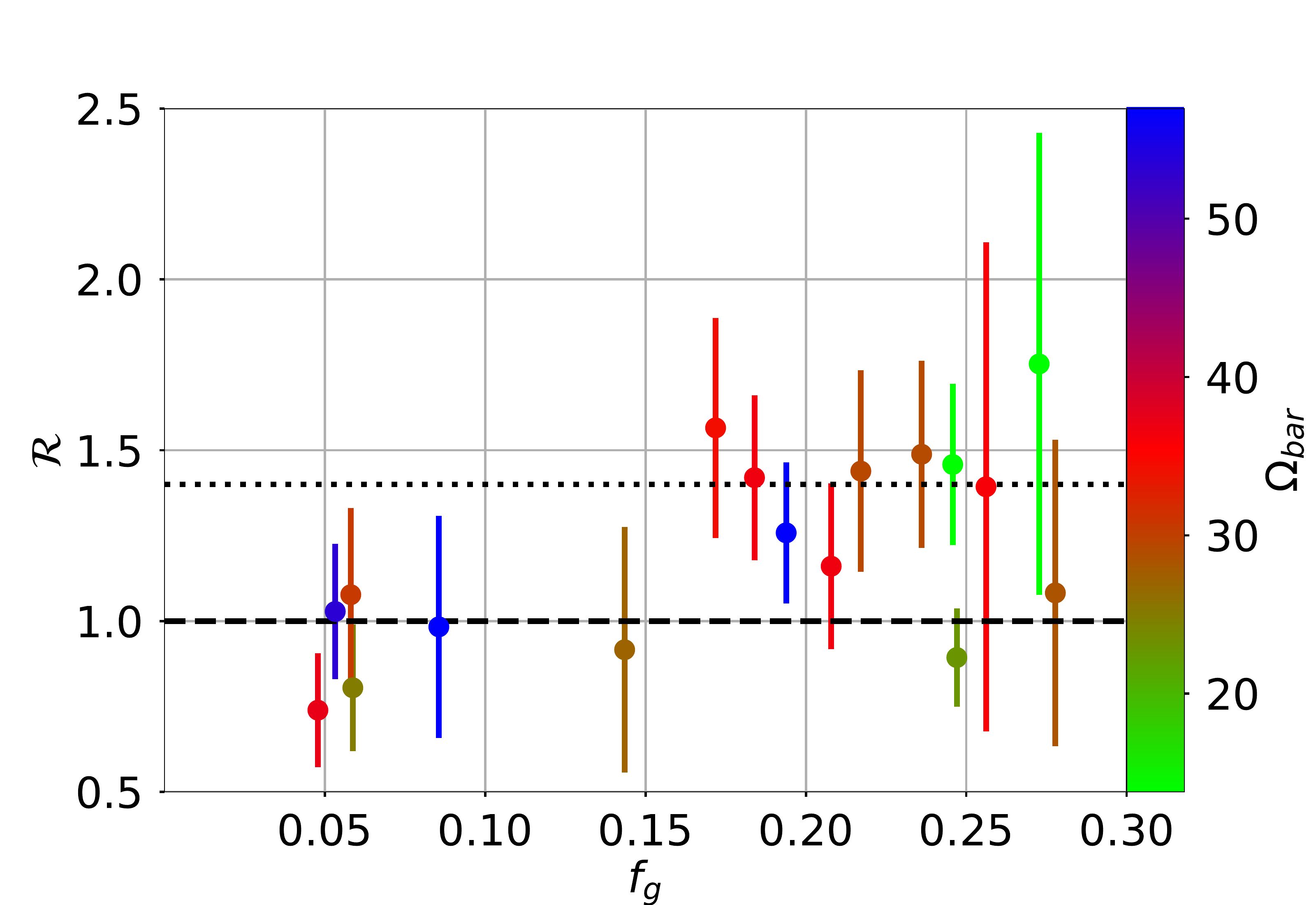}
   	 \caption{Molecular gas fraction vs. \R{} parameter. The colour code shows the median bar pattern speed. The segmented and dotted horizontal lines show the separation between ultrafast, fast and slow bars. The Spearman correlation coefficient of the relation is \Spear{0.54}}
  \label{Fig:ParR-fg}
\end{figure}

More interestingly is the relation with the molecular gas fraction $f_g = M_{gas} / (M_* + M_{gas})$. We observe a significant correlation with  \R{} and a weak anti-correlation with \Om{} (\Spear{0.54} and \Spear{-0.52} respectively). The first relation, is not reported neither in \cite{Guo2018} nor in \cite{Aguerri2015} (\Spear{0.05} in both works), however the second one it is (\Spear{-0.46} and \Spear{-0.25} respectively). In Figure \ref{Fig:ParR-fg} we show the resulting relationship between \R and the molecular gas fraction. Notice that all galaxies with $f_g < 0.15$ are consistent with the fast - ultrafast regime. This could be an indication that bars do slow down more in gaseous discs, as suggested by N-body simulations \citep{Athanassoula2013} and observations \citep{Masters2012}.

Finally we observe a weak anti-correlation between the stellar mass and \Om{} (\Spear{-0.35}) suggesting that the most massive galaxies host bigger and slower bars. This weak anti-correlation is also present in \cite{Guo2018} and \cite{Aguerri2015} (\Spear{-0.55} and \Spear{-0.30} respectively).

\section{Conclusions}
\label{Sec:Conclusions}

In this work we used the TW-method to determine \Om{} for a sample of 15 MaNGA galaxies and 3 CALIFA galaxies. To achieve this, we computed the TW-integrals over the stellar flux and stellar velocity maps produced by \code{Pipe3D}. We modelled the rotation curves using the code \code{Velfit}, taking into account the non-axisymmetric motions produced by the bar. 

We measure the disc PA using a photometric and a kinematic approach (isophotal fitting of legacy DESI images, and the best \code{Velfit} model of the central H$_\alpha$ velocity maps). To reduce the arbitrariness involved in choosing the flat PA region, we further constrained \PAph{} by using prior knowledge of \PAkn{}. This procedure proved to be useful, as the \Omph{} measurements had more physically meaningful results than \Omkn{}. Nonetheless, even with the \PAkn{} prior,  \PAph{} can be biased by external structures as illustrated by our example galaxy manga-8135-6103.

We studied error sources that arise from the geometric nature of the TW-method: the centring error (\dC{}), the PA error (\dPA{}) and the slit length error (\dL{}). We performed various tests to estimate and constrain the relative error of each source. On average, they accounted for $\sim 5 \%$, $15 \%$, $9 \%$ respectively. Each galaxy in our sample presented different sensitivities to each error. We did not observe a significant correlation between them, thus, we choose to treat them as independent error sources, and add them in quadrature for the final estimation of \Om{}. 

We tested if a lower spatial resolution of the bar could be affecting our measurements by degrading the spatial resolution in our CALIFA sub-sample. Our test showed that increasing $\sigma_{eff}$ (ie., lowering the resolution) tends to increase the sensitivity of the TW-method to the PA error, probably due to the mixing between different structures. Therefore, the resolution error should not be treated as an independent error source. We consider its effects to be already implicit in our error estimation.  Quantifying how the underlying uncertainties propagate deserves further exploration.

After constraining the uncertainties of \Om{} we used MC simulations to estimate the PDF of \CR{} and the \R{} parameter. Using the most probable bar kinematic classification, we found 7 slow, 4 fast and 7 ultrafast bars. Adding together all the distributions results in a global distribution that can be described by a double Gaussian model. 

After looking at preliminary correlations between some galactic properties and our bar measurements we observed some interesting trends. Namely a strong anti-correlation between \Om{} vs \Rbar{}, a correlation between \CR{} and \Rbar{}, a correlation between $f_g$ and \R{}, a weak-correlation between $f_g$ and \Om{} and a weak anti-correlation between \Om{} and the stellar mass. However, the strength of these relations was different in previous works \citep{Aguerri2015, Guo2018}. With the current level of uncertainties, and given the small numbers involved, these relations should be considered only exploratory in nature. Hints of the underlying bar dynamics and evolution are present, however, more accurate measurements are required.

\section{Acknowledgments}

We are grateful for the support of a CONACYT grant CB-285080. This project makes use of the MaNGA-Pipe3D dataproducts. We thank the IA-UNAM MaNGA team for creating this catalogue, and the CONACYT-180125 project for supporting them. The data processing obtained benefit from the Atocatl HPC project at IAUNAM. LGO and HHT acknowledge support from the CONACYT ‘Ciencia Basica’ grant 285721. MCD acknowledges support from UC MEXUS-CONACYT grant CN-17-128. SFS thanks CONACYT FC-2016-01-1916, and funding from the PAPIIT-DGAPA-IA101217 (UNAM) project. OV and EA acknowledge support from the UNAM grant PAPIIT-DGAPA IN112518. We also thank the referee for the helpful comments that improved the quality and interpretation of our measurements.

\bibliographystyle{mnras}
\bibliography{References}
%\input{Appendix}
%%%%%%%%%%%%%%%%%%%%%%%%%%%%%%%%%%%%%%%%%%%%%%%%%%

% Don't change these lines
\bsp	% typesetting comment
\label{lastpage}
\end{document}